\documentclass[aps,prd,twocolumn,groupedaddress,amsmath,amssymb,reprint,10 pt]{revtex4-2}

\usepackage{graphicx}
\usepackage{dcolumn}
\usepackage{bm}
\usepackage{amsmath} 
\usepackage{amssymb}
\usepackage{xcolor}
\usepackage{hyperref}
\hypersetup{
	colorlinks=true,
	linkcolor=red,
	filecolor=magenta,      
	urlcolor=blue,
	pdftitle={Selective Thermalization},
	pdfpagemode=FullScreen,
	citecolor= blue,
}
\begin{document}



\title{Four inequivalent paths to Thermality in Minkowski spacetime}


%
\author{Rakesh k Jha}
	\email{p20230070@hyderabad.bits-pilani.ac.in}
    \author{Akhil U Nair}
    \email{p20200473@hyderabad.bits-pilani.ac.in}
    \author{Prasant Samantray}
    \email{prasant.samantray@hyderabad.bits-pilani.ac.in}
   \author{Sashideep Gutti}
	\email{sashideep@hyderabad.bits-pilani.ac.in}
	\affiliation{Department of Physics, Birla Institute of Technology and Science, Pilani-Hyderabad Campus \\Hyderabad, 500078, India}	
	\date{\today}


\date{\today}

\begin{abstract}
    In this article we explore the answer to an inverse question to Unruh effect. Given a two dimensional Rindler wedge $R$ with a thermal distribution of massless scalar particles, we explore the supersets of $R$, whose reduced state yields the observed particle content in $R$. If we restrict our analysis to particle content close to the Rindler horizon in $R$, we show that the answer to the inverse question is not unique. The analysis is done using two methods, first uses Bogolibov coefficients and the second method uses Virasoro anomaly. We identify four inequivalent paths with various "parent" spacetimes of $R$ from which we can arrive at the given particle content in $R$. We show that the supersets are: Minkowski spacetime, a Rindler wedge in vacuum, a Rindler wedge with a thermal flux of left moving particles and lastly a Rindler wedge with a thermal flux of right moving flux of particles. We also show an interesting phenomenon: flux to density conversion, wherein a thermal flux of either left moving or right moving particles can be converted to thermal density of particles involving both left moving and right moving particles. Based on the analysis done in the article, we present qualitative arguments regarding the possibility that the continuous evaporation of a black hole might be punctuated by a series of pauses and bursts of radiation.  
\end{abstract}


\maketitle{}

\section{Introduction \label{Sec-1}}
The Unruh Davies effect is a well-studied and well-understood phenomenon. When we consider a two-dimensional Minkowski spacetime with a quantum field in its vacuum state, a uniformly accelerating observer perceives the field in vacuum as an excited state with a thermal distribution of particles~\cite{Unruh:1976db}. A Rindler spacetime is a static spacetime constructed by considering a continuum of accelerating observers. It is a subset of Minkowski spacetime, and when a Minkowski spacetime contains a quantum field in a vacuum state, it appears as a thermal state in Rindler spacetime. A related idea is that of Hawking radiation. An astrophysical black hole is predicted to radiate a thermal flux of radiation known as Hawking radiation with a temperature roughly proportional to the inverse of black hole mass~\cite{Hawking:1975vcx}. Rindler spacetime offers a simpler setting in which black hole radiation can be studied and understood. 

Nested and shifted Rindler wedges provide a rigorous geometric framework for exploring observer-dependent thermodynamics and subsystem entanglement in quantum field theory. Exploring the kinematic consequences of nesting, Lochan and Padmanabhan showed that sequences of spatially shifted wedges exhibit ``relic thermality": the vacuum of an enveloping wedge appears universally thermal to any strictly nested observer, regardless of the spatial shift's magnitude~\cite{Lochan:2025mru}. A shifted Rindler model was also used in Ref.~\cite{Gutti:2022xov}, who demonstrated that tracing out degrees of freedom across nested regions (e.g., $R_2 \subset R_1$) can be path-dependent. Furthermore, extensions to null-shifted wedges by the authors in Ref.~\cite{Jha:2025tpg} revealed ``selective particle excitations," a geometric consequence where only specific momentum modes or chiralities are thermally excited. Ultimately, these works demonstrate that the precise causal nesting of Rindler wedges can be used as toy models to study dynamical evolution of black hole apparent horizon. 
\begin{figure}[!ht]
	\centering
	\includegraphics[scale=0.6]{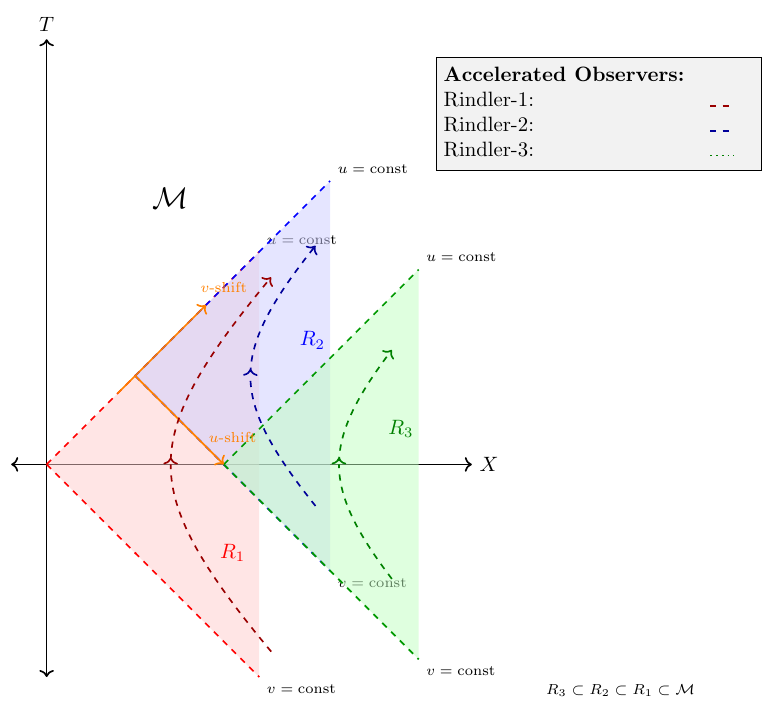}
	\caption{Null-shifted Rindler wedges in Minkowski spacetime. The standard right Rindler wedge is successively shifted along the future-directed null $V$-direction to obtain  ($R_2$) and then along the $U$-direction to obtain ($R_3$).
    \label{Fig:1}}
\end{figure}
In this article, we are interested in the converse question to the Unruh effect. We have a Rindler wedge $R_3$ with a massless scalar field with particle distribution in a thermal state. What are the supersets $\mathcal{A}$ of $R_3$ ($R \subset A$) whose reduced state yields a thermal state in $R_3$?  Stated differently, if we have a uniformly accelerating observer in Minkowski spacetime who detects a thermal bath of scalar particles, what can he/she deduce about the spacetime beyond his/her horizon? Since he/she is a uniformly accelerating observer, he/she has a future and past horizon, and he/she is therefore restricted to a Rindler wedge. We ask a converse question. What are the possible supersets of his/her Rindler wedge whose reduced state yields the observed particle distribution? One obvious answer is, of course, Minkowski spacetime with the scalar field in vacuum. But is the answer unique? We justify in this paper that the answer is not unique by showing that thermality in a given Rindler wedge can occur along four inequivalent paths; therefore, we argue that the converse question has multiple answers.\\
Path one is as stated before, is from a Minkowski spacetime with the scalar field in a vacuum state. Path two involves a Rindler wedge $R_1$ with the scalar field in vacuum (the Rindler wedge $R_1$ is located such that $R_3 \subset R_1$, such that the bifurcation point of $R_1$ and $R_3$ are separated by a spacelike interval. Path three starts with a Rindler wedge $R_2$  (with $R_3 \subset R_2$) with an inward flux of scalar particles. $R_3$ is a wedge shifted from $R_2$ along a null ray as explained in later sections.
Path four starts with a Rindler wedge $R_4$ again (with $R_3 \subset R_4$), and in contrast with path three, there is an outward thermal flux of scalar particles. $R_3$ is a wedge shifted along a null ray from $R_4$ as illustrated in later sections.

We also show a new interesting effect: flux-to-density conversion. We consider a Rindler wedge with either an outgoing or an ingoing particle flux and show how to convert it to a particle density in a Rindler wedge that is a subset of the previous wedge. The paths three and four in the previous paragraph illustrate this effect.

We mention a few aspects and assumptions followed in this work. The analysis is done using two methods. One is via mode functions and the explicit evaluation of the Bogoliubov coefficients. The Bogoliubov transformation approach operates at the level of mode expansions, mixing creation and annihilation operators to relate distinct vacua and thereby connect asymptotic particle states. The other method uses the Virasoro anomaly in two-dimensional conformal field theory (2D CFT). Since we consider a massless scalar field in 2D for this work, the techniques of 2D conformal field theory are applicable. We use the Virasoro anomaly to evaluate the stress tensor and thereby determine the particle content in various Rindler wedges. By contrast to the Bogoliubov approach, the stress tensor and conformal anomaly method evaluate renormalized local operator expectation values, such as energy fluxes, and reveal how anomalies encode radiation and particle production. While the Bogoliubov method emphasizes the relation between asymptotic states, the stress tensor method provides a local field‑theoretic account of fluxes and anomalies. Both approaches ultimately describe the same physical phenomenon, but from different calculational perspectives. In the analysis presented in this article, we take the assumption that the frequencies involved are large and that the calculations are applicable near the horizon of the uniformly accelerating observer. 

As mentioned before, the starting point of Path-2 is the Rindler spacetime with a scalar field in the Rindler vacuum. Now, the Rindler vacuum is not a physically well-defined state in the context of Minkowski spacetime, since the local stress tensor with respect to the Rindler vacuum diverges on the horizon. As demonstrated in past literature~\cite{Aalsma:2019rpt}, the Boulware vacuum (in the near-horizon limit, the Boulware vacuum behaves like the Rindler vacuum, as the Boulware vacuum is defined in curved spacetime and globally differs from the Rindler vacuum, whereas the Rindler vacuum lives in flat spacetime and only covers part of it) in the context of black holes.
The Rindler spacetime and the analysis done in the article serve as a toy model for the realistic case involving evolving black holes and the possible signatures of the evolving apparent horizons. As pointed out earlier, the Rindler spacetime maps to the Boulware vacuum in curved spacetime. As we discuss in the conclusions section of the article, the Rindler vacuum in the toy model and the corresponding Boulware vacuum in the curved spacetime setting may play a role in the context of black hole evaporation.



\section{The Problem Statement and SetUp \label{Sec-2}}

Suppose we have a Rindler wedge $R_3$ in which we consider a real massless scalar field with the particle content in a thermal distribution. We are interested in the supersets $A$ of $R_3$ that yield the observed particle content to be thermal upon finding the reduced state in $R_3$ from $A$. As stated in the introduction, we show that $A$ is not unique. We begin by formulating the setup necessary for the analysis and subsequently present the different possible paths.

\subsection{Four inequivalent paths}
We discuss below the various paths leading to a common endpoint, viz., the wedge $R_3$ with a thermal distribution of particles in the high-acceleration regime (close to the horizons). As stated before in the introduction, we show that various starting points yield the common endpoint. By `starting points,' we mean various spacetimes with massless scalar fields in their corresponding particle states. We demonstrate this by explicitly choosing the starting points and evaluating the final particle content in $R_3$ along the given paths. The coordinate definitions are given below,
Two dimensional Minkowski spacetime is denoted as $\mathcal{M}$, parametrized by inertial coordinates $(X, T)$, and a hierarchy of Rindler wedges~\cite{Rindler:1960zz}, denoted $(R_i)$, each described by coordinates $(x_i, t_i)$ for $i \in \{1,2,3,4\}$.  As depicted in Figs .~\ref{Fig:1} and ~\ref{Fig:2}, the nested structure of Rindler wedges are understood as inclusions of spacetime regions determined by their null horizons, such that $R_3 \subset R_2 \subset R_1 \subset M$, and $R_3 \subset R_4 \subset R_1 \subset M$, with individual wedges distinguished by the shaded regions in pale red, blue, and green. Wedges $R_2$ and $R_3$ are generated from $R_1$ by successive null displacements along the relevant Minkowski light rays, a construction central to our analysis.
\subsubsection{Path 1: Minkowski (in vacuum state) \texorpdfstring{$M$}{M} to Rindler wedge \texorpdfstring{$\mathrm{R_3}$}{R3}}
In this path, we consider Minkowski spacetime with the real massless scalar field in a vacuum state. The wedge $R_3$ is located such that its bifurcation point is situated to the right of the origin. The Unruh effect states that the reduced state of the massless scalar field in $R_3$ is a thermal distribution of particles. Keeping in perspective the method followed in the article, we note that we can also arrive at this result by subdividing this path into two sub-steps. We note that the metric in coordinates $T, X$ is $ds^2 = -dT^2 + dX^2$. We choose null rays as $U=T-X$ and $V=T+X$. The two sub-paths are stated below: a) We divide the Minkowski spacetime into two parts by choosing coordinate patches $U, V$ with a new map $u,V$ with $u$ given by $U=-e^{-au}/a-\delta$  where $a$ is an arbitrary real positive parameter. We note that the null coordinates range over $-\infty<U<\infty$ and, similarly, $-\infty<u<\infty$. We note that when coordinate $u$ goes from $-\infty$ to $\infty$, the coordinate $U$ ranges from $0$ to $\infty$. This is defined as half-sided modular inclusion~\cite{Wiesbrock:1992mg}. b) We now go from half of Minkowski spacetime to one fourth by defining one more coordinate transformation given by $u,v$ where $u$ is the same as in subpart a and $V=e^{av}/a+\gamma$ with $\gamma$ a positive constant. We can see that this gives us the Rindler right wedge, provided we note the restriction that $U>0$ and $V<0$. We note that we could interchange the order of subparts $a$ and $b$. In the standard Unruh effect, it is well known that this ordering does not make a difference to the final particle content in the Rindler wedge.
\subsubsection{Path 2: Rindler wedge \texorpdfstring{$R_1$}{R1} in vacuum state to Rindler wedge \texorpdfstring{$R_3$}{R3} via Spatial Shift}
In Path 2, the wedge $R_3$ is defined from $R_1$ by a spatial translation, characterized by displacement parameter $\Delta_1$ (cf. Figure~\ref{Fig:1}). Although the bifurcation point of $R_3$ is shown along the $x_1$ axis, its precise location may be generalized to any point within the $R_1$ wedge. Though the answer to this path is worked out in Refs.~\cite{Lochan:2025mru, Gutti:2022xov} using the method of Bogoliubov coefficients, we present an alternative analysis here using the Virasoro anomaly approach. We note that in a given Rindler wedge, say for example in $R_1$, the magnitude of the acceleration of an observer progressively decreases as one moves to the right along the $t_1=0$ line ($x_1-axis$). This acceleration tends to zero as one approaches $ x_1->\infty$. The same is true for an observer in $R_3$ as one moves to the right ($x_3->\infty$). It is reasonable to expect that any particle excitations in the wedge $R_3$ are more prominent close to the horizons. We demonstrate this in the article. 

\subsubsection{Path 3: Null Shifts}
Consider a Rindler wedge $R_2$ with the following particle content. The left-moving particles are present with a thermal distribution, while there are no right-moving particles. There is therefore a net left-moving thermal flux of particles. We show that starting from $R_2$ with the above-mentioned particle distribution, we arrive at a thermal distribution of particles in $R_3$ with both the left-moving and right-moving particles with a thermal distribution of particles with the standard Rindler temperature. We note that this is true close to the horizon.
   
An alternative definition of path 3 traces a sequence $R_1 \rightarrow R_2 \rightarrow R_3$, where $R_2$ is obtained from $R_1$ by displacing its origin along the future-directed null $V_1$-axis by a Minkowski parameter $\Delta_2$,
$R_3$ is then constructed by a further displacement of $R_2$ along the null $U_2$-axis by parameter $\Delta_3$ (see Figure~\ref{Fig:1}).
For illustration, Figure~\ref{Fig:1} depicts the case where $\Delta_2 = \Delta_3$, but this equality is not a physical constraint and can be ensured by an appropriate Lorentz transformation, since these parameters are not Lorentz scalars. In our article, we present the latter definition of path 3, where the starting point is again a vacuum in wedge $R_1$.

\subsubsection{Path 4: Null Shifts via Rindler Wedge \texorpdfstring{ $R_4$}{R4}} 
Path 4 involves the wedge \(R_4\), whose origin is obtained from \(R_1\) by a null shift along the \(U_1\)-axis by parameter \(\Delta_4\). Subsequently, \(R_3\) is reached by an additional null shift from \(R_4\) along the \(V_2\)-axis with parameter \(\Delta_3\), as depicted in Figure~\ref{Fig:2}. The wedge \(R_4\) is an independent Rindler wedge analogous to \(R_2\), obtained by shifting through a different null axis. The idea is the same as that of path 3 except that wedge $R_2$ is replaced with wedge $R_4$. Again, we note that the starting point of path 4 could be that the wedge $R_4$ has a right-moving thermal flux of radiation, whereas there is no particle content in the right-moving sector. For this initial configuration as well, the particle distribution in $R_3$ is thermal. Although this is possible, in this work, we focus on an analysis that begins with the Rindler wedge $R_1$ in the vacuum state.

\subsection{Coordinate Relations among the key players \label{Subsec-2.1}}
In this subsection, we introduce the various coordinate systems relevant to our setup and describe the relations among them.
\subsubsection{ Minkowski spacetime to Rindler $1$ wedge}
We label the coordinates of the Rindler-1 ($R_1$) frame as \((x_1, t_1)\).
The transformation between the Rindler-1 coordinates ($R_1$) and Minkowski ($M$) is given by:
\begin{equation}
	T = \frac{e^{a_1 x_1}}{a_1} \sinh(a_1 t_1), \label{Eq:2.1.0.1}
\end{equation}

\begin{equation}
	X = \frac{e^{a_1 x_1}}{a_1} \cosh(a_1 t_1) ,\label{Eq:2.1.0.2}
\end{equation}
we now introduce the light-cone coordinates  in $R_1$, $(u_1,v_1)$:
\begin{equation}
    u_1 = t_1 - x_1, \quad v_1 = t_1 + x_1, 
    \label{Eq:2.1.0.2.a}
\end{equation}
And the lightcone coordinates in Minkowski spacetime are related to the coordinates of $R_1$ as:
\begin{equation}
    U_M = T-X = -\frac{e^{-a_1 u_1}}{a_1}, 
    \label{Eq:2.1.0.3}
\end{equation}
\begin{equation}
    V_M = T+X = \frac{e^{a_1 v_1}}{a_1}, 
    \label{Eq:2.1.0.4}
\end{equation}

\subsubsection{Minkowski to Rindler-3 and Rindler-1 to Rindler-3 via spatial shift along  \texorpdfstring{$x_1$}{x1} axis}
We label the coordinates of the Rindler-3 ($R_3$) as \((x_3, t_3)\). In the setup, we already mentioned that we do not assume any intermediate wedges between $R_1$ and $R_3$. We now relate Rindler-3 ($R_3$) to Minkowski space. Its transformation is given by:
\begin{align}
	T &= \frac{e^{a_3\; x_3}}{a_3} \sinh(a_3\; t_3) , \label{Eq:2.1.0.5} \\
	X &= \frac{e^{a_3\; x_3}}{a_3} \cosh(a_3\; t_3) + \Delta_1 . \label{Eq:2.1.0.6}
\end{align}
where $\Delta_1$ encodes the coordinate shift in the Minkowski spacetime, along $x_{1}$ axis relates the wedge $R_3$ to $R_1$ ,if possible we can shown $\Delta_1$ in figure.
The lightcone coordinates in Minkowski spacetime are related to the coordinates of $R_3$ as:
\begin{align}
    U_M = T-X &= -\frac{e^{-a_3 \;u_3}}{a_3}-\Delta_1, 
    \label{Eq:2.1.0.7}
\\
    V_M = T+X &= \frac{e^{a_3\; v_3}}{a_3}+\Delta_1, 
    \label{Eq:2.1.0.8}
\end{align}
In a given Rindler frame, the parameter that represents acceleration is $a_1$ for the wedge $R_1$. We note that if we choose an alternative parameter $b_1$ as the acceleration parameter, we obtain a different coordinate system for the same wedge. One can set up the description of the quantum field in the new coordinate system. It is easy to prove that both descriptions share the vacuum state, as the Bogoliubov coefficients between the modes defined in either coordinate system mix only positive-frequency modes of one coordinate system with the positive-frequency modes of the other coordinate system. Therefore, we are free to choose whichever parameter we want in a given Rindler wedge. This choice amounts to selecting a particular accelerating observer from among a family of accelerating observers that accelerate at different proper accelerations, all of which share the same horizon. Therefore, we set $ a_1 = a_2 = a_3 = a$, which simplifies the calculation without losing generality.
Equating  Eq.~(\ref{Eq:2.1.0.3}) and  Eq.~(\ref{Eq:2.1.0.7}), we have;
\begin{equation}
  u_{1}  = -\frac{1}{a}ln(e^{-a\;u_{3}} + a\;\Delta_1), \label{Eq:2.1.0.9}
\end{equation}
Also from   Eq.~(\ref{Eq:2.1.0.4}) and  Eq.~(\ref{Eq:2.1.0.8}), we have;
\begin{equation}
   v_{1} = \frac{1}{a}ln(e^{a \;v_3}+ a\; \Delta_{1}) ,\label{Eq:2.1.0.10}
\end{equation}
From Eq.~(\ref{Eq:2.1.0.9}) and  Eq.~(\ref{Eq:2.1.0.10}) and at $\Delta_{1}=\frac{1}{a}$ we obtain,
\begin{equation}
   u_{1} = -\frac{1}{a}ln(e^{-a\;u_{3}} + 1), \label{Eq:2.1.0.11}
   \end{equation}
   \begin{equation}
   v_{1} = \frac{1}{a}ln(e^{a \;v_3}+ 1).\label{Eq:2.1.0.12}
\end{equation}

\subsubsection{Rindler-1 to Rindler-3 via Rindler-2:  series of null shifts}
We label the coordinates of the Rindler-1 ($R_1$) frame as \((x_1, t_1)\), Rindler-2 ($R_2$) as \((x_2, t_2)\), and Rindler-3 ($R_3$) as \((x_3, t_3)\).
\begin{figure}[!ht]
	\centering
	\includegraphics[scale=0.6]{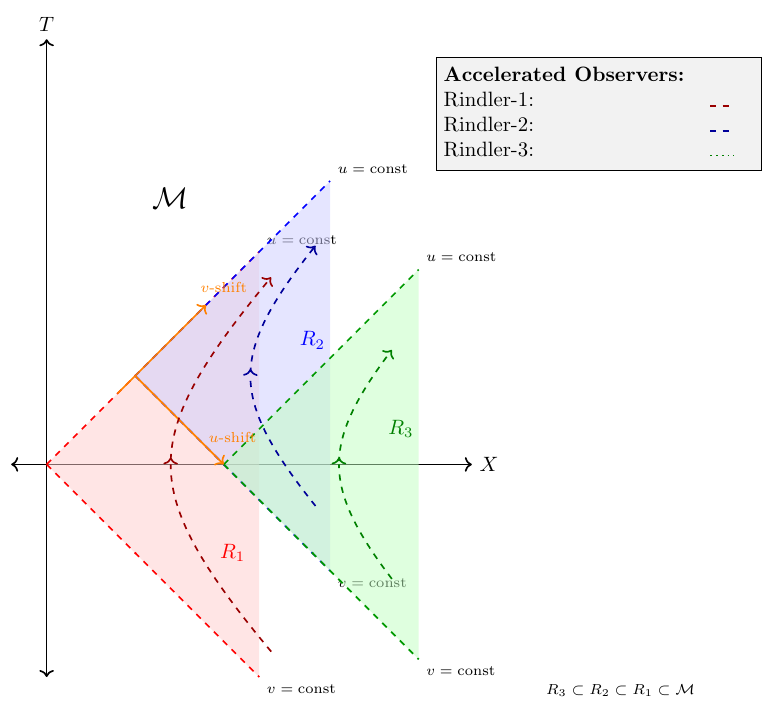}
	\caption{Null-shifted Rindler wedges in Minkowski spacetime. The standard right Rindler wedge is successively shifted along the future-directed null $V$-direction to obtain  ($R_2$) and then along the $U$-direction to obtain ($R_3$).
    \label{Fig:2}}
\end{figure}

The frame $R_2$ is null shifted along the \(V_1\) axis. Its coordinates relate to Minkowski as follows:
\begin{align}
	T &= \frac{e^{a \;x_2}}{a} \sinh(a\; t_2) + \Delta_2, \label{Eq:2.1.1.1} \\
	X &= \frac{e^{a\; x_2}}{a} \cosh(a\; t_2) + \Delta_2.\label{Eq:2.1.1.2}
\end{align}
Where $\Delta_2$ encodes the coordinate shift in the Rindler space that relates the wedge $R_2$ to $R_1$.
The lightcone coordinates of Minkowski spacetime and  those of  $R_2$ are related by:
\begin{align}
	U_M &= T - X = -\frac{e^{-a\; u_2}}{a}, \label{Eq:2.1.1.3} \\
	V_M &= T + X = \frac{e^{a\; v_2}}{a} + 2 \;\Delta_2.\label{Eq:2.1.1.4}
\end{align}
Equating  Eq.~(\ref{Eq:2.1.0.3}) and  Eq.~(\ref{Eq:2.1.1.3}), we have;
\begin{equation}
  u_{1}  = u_{2}, \label{Eq:2.1.1.5}
\end{equation}
Also from   Eq.~(\ref{Eq:2.1.0.4}) and  Eq.~(\ref{Eq:2.1.1.4}), at $\Delta_2=1/2a$ we have;
\begin{equation}
   v_1 = \frac{ln(e^{a\; v_2}+ 1)}{a} ,\label{Eq:2.1.1.6}
\end{equation}
We now relate the Rindler-3 ($R_3$) coordinate system to Minkowski spacetime coordinates. Its transformations are given by:
\begin{align}
	T &= \frac{e^{a\; x_3}}{a} \sinh(a\; t_3) + (\Delta_2 - \Delta_3), \label{Eq:2.1.1.7} \\
	X &= \frac{e^{a\; x_3}}{a} \cosh(a\; t_3) + (\Delta_2 + \Delta_3). \label{Eq:2.1.1.8}
\end{align}
where $\Delta_3$ encodes the coordinate shift in the Rindler space that relates the wedge $R_3$ to $R_2$.
In light-cone form:
\begin{align}
	U_M &= T - X = -\frac{e^{-a\; u_3}}{a} - 2\; \Delta_3, \label{Eq:2.1.1.9} \\
	V_M &= T + X = \frac{e^{a\; v_3}}{a} + 2\; \Delta_2.\label{Eq:2.1.1.10}
\end{align}
Matching with the coordinates of $R_2$, at\;$\Delta_2=1/2a$ we find:
\begin{align}
	u_2 &= -\frac{ln[e^{-a \;u_3}+ 1]}{a}, \label{Eq:2.1.1.11} \\
	v_2 &= v_3. \label{Eq:2.1.1.12}
\end{align}
 Without loss of generality, we choose $\Delta_3=\Delta_2$. This is because we can choose a suitable Lorentz transformation in which both the parameters are equal, and hence the bifurcation point of $R_3$ lies on the $x_1$ axis, simplifying the geometry without altering the physical content.

\subsubsection{Rindler-1 to Rindler-3 via Rindler-4:  series of  null shifts}
Analogously, consider null shifts along the \( U \)-axis, as shown in Fig.~\ref{Fig:2}
\begin{figure}[!ht]
	\centering
	\includegraphics[scale=0.6]{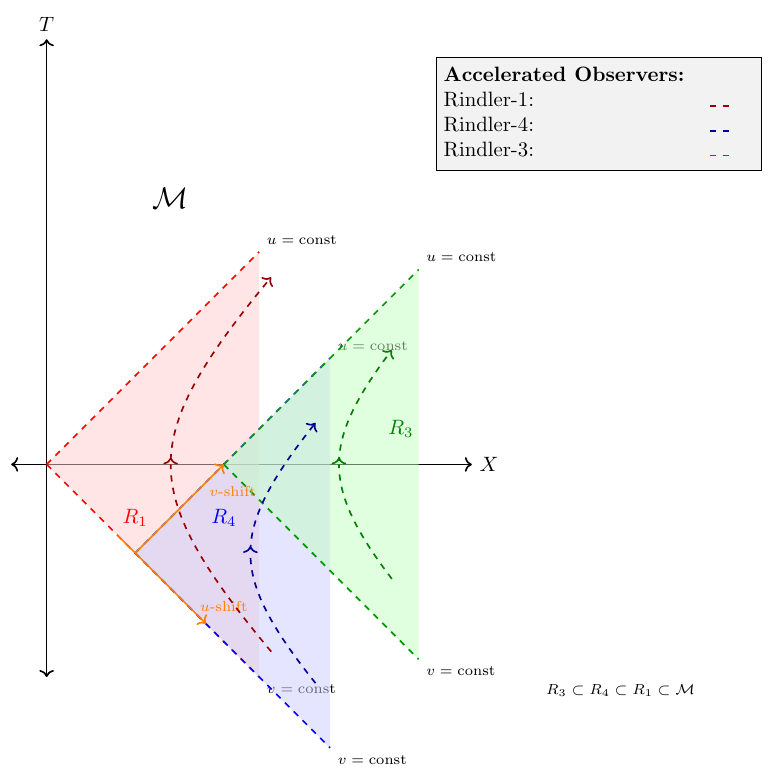}
	\caption{Alternative null-shift construction. The wedge ($R_4$) is obtained from ($R_1$) by a null shift along the $ U$-direction, followed by a shift along the  $V$-direction to obtain ($R_3$). As in Fig.~\ref{Fig:1}, the construction results in nested spacetime regions.  \label{Fig:3}}
\end{figure}

The frame $R_1$ remains as:
\begin{align}
	T &= \frac{e^{a\; x_1}}{a} \sinh(a \;t_1), \label{Eq:2.1.1.13} \\
	X &= \frac{e^{a\; x_1}}{a} \cosh(a\; t_1). \label{Eq:2.1.1.14}
\end{align}
In light-cone form:
\begin{align}
	U_M &= -\frac{e^{-a\; u_1}}{a}, \label{Eq:2.1.1.15} \\
	V_M &= \frac{e^{a\; v_1}}{a}. \label{Eq:2.1.1.16}
\end{align}

Now consider $R_4$, shifted along the \( U \)-axis.
\begin{align}
	T &= \frac{e^{a \;x_4}}{a} \sinh(a\; t_4) - \Delta_4, \label{Eq:2.1.1.17} \\
	X &= \frac{e^{a\; x_4}}{a} \cosh(a\; t_4) + \Delta_4. \label{Eq:2.1.1.18}
\end{align}
In light-cone form:
\begin{align}
	U_M &= -\frac{e^{-a\; u_4}}{a} - 2 \;\Delta_4, \label{Eq:2.1.1.19} \\
    V_M &= \frac{e^{a\; v_4}}{a}. \label{Eq:2.1.1.20}
\end{align}

From Eq.~(\ref{Eq:2.1.1.15}) and Eq.~(\ref{Eq:2.1.1.19}), at $\Delta_4=1/2a$ we obtain:
\begin{equation}
	u_1 = -\frac{ln(e^{-a\;u_4}+1)}{a} , \label{Eq:2.1.1.21}
\end{equation}
and from Eq.~(\ref{Eq:2.1.1.16}) and Eq.~(\ref{Eq:2.1.1.20}):
\begin{equation}
	v_1 = v_4, \label{Eq:2.1.1.22}
\end{equation}

Finally, for $R_3$:
\begin{align}
	T &= \frac{e^{a\; x_3}}{a} \sinh(a\; t_3) - (\Delta_4 - \Delta_3), \label{Eq:2.1.1.23} \\
	X &= \frac{e^{a\; x_3}}{a} \cosh(a \;t_3) + (\Delta_4 + \Delta_3). \label{Eq:2.1.1.24}
\end{align}
In light-cone form:
\begin{align}
	U_M &= -\frac{e^{-a \;u_3}}{a} - 2\; \Delta_4, \label{Eq:2.1.1.25} \\
	V_M &= \frac{e^{a\; v_3}}{a} + 2 \;\Delta_3. \label{Eq:2.1.1.26}
\end{align}

Matching the coordinates of $R_4$, and also without loss of generality, we choose $\Delta_4 = \Delta_3 = \frac {1}{2a}$. We then obtain the following.
\begin{align}
	u_4 &= u_3, \label{Eq:2.1.1.27} \\
	v_4 &= \frac{1}{a} \ln\left(e^{a\; v_3} + 1\right).\label{Eq:2.1.1.28}
\end{align}
\noindent
These expressions completely characterize the relationships among the Rindler patches that shift to the null in the $U$ and $V$ directions.

These construction paths serve to systematically characterize inequivalent purifications of Rindler vacua and reveal how selective and full thermalization arise in distinct regions of spacetime. While intermediate wedges are omitted in some scenarios for simplicity, the analysis extends to more general initial states, though with potentially differing particle spectra~\cite{Gutti:2022xov}.

\section{Methodology}
In the previous section, we defined the coordinate systems in various spacetimes of relevance and obtained the relations between them. To determine the particle content in various spacetimes, we use two techniques. One is the canonical way to set up the mode expansion of scalar fields and to evaluate the Bogoliubov coefficient. Another useful method is using the Virasoro anomaly.

\subsection{Bogoliubov method}
We begin by considering a free, massless scalar field in $(1+1)$-dimensional Rindler spacetime. The field satisfies the Klein–Gordon equation,
\begin{equation}
	\square \,\hat{\phi}(x^\mu) = 0,
	\label{Eq:3.0.0.1}
\end{equation}
where $\square = \eta^{\mu\nu} \nabla_\mu \nabla_\nu$.  

In conformally flat form, the Rindler metric allows Eq.~(\ref{Eq:3.0.0.1}) to be rewritten as
\begin{equation}
    e^{2 a x}\square \,\hat{\phi}(x^\mu) 
    = \left(\frac{\partial^2}{\partial t^2}-\frac{\partial^2}{\partial x^2}\right)\hat{\phi}(x^\mu)= 0.
	\label{Eq:3.0.0.2}
\end{equation}
In $(1+1)$ dimensions, the general solution separates into right-moving and left-moving modes, propagating along the null coordinates $u = t-x, \qquad v = t + x$.
Accordingly, the quantum scalar field operator in a single Rindler wedge $R_i$ can be expanded as~\cite{Wald:1984rg}
\begin{equation}
\begin{split}
	\hat{\phi}(x_i,t_i) =
	\int_{0}^{\infty} d\omega \bigg[ & \overrightarrow{\hat{a}}_i(\omega)\; f_i(\omega) +
    \overrightarrow{\hat{a}}_i^\dagger(\omega)\; f^*_i(\omega)  \\
    &+ \overleftarrow{\hat{b}}_i(\omega)\; g_i(\omega) +
    \overleftarrow{\hat{b}}_i^\dagger(\omega)\; g^*_i(\omega)\bigg],
\end{split}
    \label{Eq:3.0.0.3}
\end{equation}
where $\overrightarrow{ \hat{a}}_i(\omega)$ and $\overleftarrow{ \hat{b}}_i(\omega)$ are the annihilation operators associated with right- and left-moving modes, respectively. The mode functions take the form
\begin{equation*}
     f_i(\omega) = N_\omega \, e^{-i\omega(t-x)},
     \qquad
     g_i(\omega) = N_\omega \, e^{-i\omega(t+x)}.
\end{equation*}

The normalization constant $N_\omega$ is determined by requiring orthonormality of the modes under the Klein–Gordon inner product, defined on a constant-time hypersurface $\Sigma$ as~\cite{Wald1993, Frodden:2018mdm}
\begin{equation}
	\langle \phi_1,\phi_2 \rangle = i \int_{\Sigma} d^{\,n-1}\Sigma \; \sqrt{\gamma}\;\eta^\mu 
	\left( \phi_1^* \nabla_\mu \phi_2 - \phi_2\nabla_\mu \phi_1^* \right).
	\label{Eq:3.0.0.4}
\end{equation}
For right- and left-moving modes, this reduces to~\cite{fabbri12005}
\begin{align}
    \langle f_l,f_m \rangle &= -i \int_{\Sigma} du \, \left( f_l \,\partial_u f^*_m - f^*_m\,\partial_u f_l \right), \label{Eq:3.0.0.5}\\
    \langle g_l,g_m \rangle &= -i \int_{\Sigma} dv \, \left( g_l \,\partial_v g^*_m - g^*_m\,\partial_v g_l \right),
	\label{Eq:3.0.0.6}
\end{align}
for $l \neq m = 1,2$. From this normalization condition, one obtains
\begin{equation}
    N_\omega = \frac{1}{\sqrt{4\pi \omega}}.
	\label{Eq:3.0.0.7}
\end{equation}

It follows that the full scalar field operator splits into independent right- and left-moving sectors,
\[
\hat{\phi}(x) = \hat{\phi}(u) + \hat{\phi}(v).
\]
The right-moving component in wedge $R_1$ takes the form
\begin{equation}
    \hat{\phi}_{1 }(u_1)= \int_{0}^\infty \frac{d\omega}{\sqrt{4\pi \omega}} 
    \left[\overrightarrow{\hat{a}}_{1}(\omega)\, e^{-i\omega u_{1}}
    + \overrightarrow{\hat{a}}_{1}^\dagger(\omega)\, e^{i\omega u_{1}}\right],
    \label{Eq:3.0.0.8}
\end{equation}
while in the remaining wedges, the expansions are
\begin{align}
    \hat{\phi}_{2}(u_2) &= \int_0^\infty \frac{d\Omega}{\sqrt{4\pi \Omega}} 
    \left[ \overrightarrow{\hat{a}}_{2}(\Omega)\, e^{-i\Omega u_{2}} 
    + \overrightarrow{\hat{a}}_{2}^\dagger(\Omega)\, e^{i\Omega u_{2}} \right], \label{Eq:3.0.0.9}\\
    \hat{\phi}_{3}(u_3) &= \int_0^\infty \frac{d\nu}{\sqrt{4\pi \nu}} 
    \left[ \overrightarrow{\hat{a}}_{3}(\nu)\, e^{-i\nu u_{3}} 
    + \overrightarrow{\hat{a}}_{3}^\dagger(\nu)\, e^{i\nu u_{3}} \right], \label{Eq:3.0.0.10}\\
    \hat{\phi}_{4}(u_4) &= \int_0^\infty \frac{d\rho}{\sqrt{4\pi \rho}} 
    \left[ \overrightarrow{\hat{a}}_{4}(\rho)\, e^{-i\rho u_{4}} 
    + \overrightarrow{\hat{a}}_{4}^\dagger(\rho)\, e^{i\rho u_{4}} \right].
    \label{Eq:3.0.0.11}
\end{align}

Analogously, the left-moving components of the field operator are
\begin{align}
    \hat{\phi}_{1}(v_1) &= \int_0^\infty \frac{d\omega}{\sqrt{4\pi \omega}} 
    \left[ \overleftarrow{\hat{b}}_{1}(\omega)\, e^{-i\omega v_{1}} 
    + \overleftarrow{\hat{b}}_{1}^\dagger(\omega)\, e^{i\omega v_{1}} \right], \label{Eq:3.0.0.12}\\
    \hat{\phi}_{2}(v_2) &= \int_0^\infty \frac{d\Omega}{\sqrt{4\pi \Omega}} 
    \left[\overleftarrow{\hat{b}}_{2}(\Omega)\, e^{-i\Omega v_{2}} 
    + \overleftarrow{\hat{b}}_{2}^\dagger(\Omega)\, e^{i\Omega v_{2}} \right], \label{Eq:3.0.0.13}\\
    \hat{\phi}_{3}(v_3) &= \int_0^\infty \frac{d\nu}{\sqrt{4\pi \nu}} 
    \left[ \overleftarrow{\hat{b}}_{3}(\nu)\, e^{-i\nu v_{3}} 
    + \overleftarrow{\hat{b}}_{3}^\dagger(\nu)\, e^{i\nu v_{3}} \right], \label{Eq:3.0.0.14}\\
    \hat{\phi}_{4}(v_4) &= \int_0^\infty \frac{d\rho}{\sqrt{4\pi \rho}} 
    \left[\overleftarrow{ \hat{b}}_{4}(\rho)\, e^{-i\rho v_{4}} 
    + \overleftarrow{\hat{b}}_{4}^\dagger(\rho)\, e^{i\rho v_{4}} \right].
    \label{Eq:3.0.0.15}
\end{align}

Here, the frequency parameters $\omega, \Omega, \nu,$ and $\rho$ denote the mode labels associated with wedges $R_1$, $R_2$, $R_3$, and $R_4$, respectively.
\subsubsection{Bogoliubov Transformation and the Number Operator \label{Subsec-3.1}} 
In this section, we derive the Bogoliubov transformation relating two complete sets of mode functions and use it to compute the expectation value of the number operator in different Rindler frames.

Let \(\{f_i(\omega)\}\) and \(\{f_j(\Omega)\}\) denote two complete sets of positive-norm, orthonormal mode functions with respect to the Klein–Gordon inner product defined in Eq.~(\ref{Eq:3.0.0.4}). These mode sets are related by the Bogoliubov transformation~\cite{Sriramkumar:1999nw, Birrell:1982ix}:
\begin{equation}
  f_j(\Omega) = \int_{-\infty}^\infty d\omega \left[
    \alpha_{ji}(\Omega,\omega) f_i(\omega) + \beta_{ji}(\Omega,\omega) f_i^*(\omega)
  \right],
  \label{Eq:3.1.0.1}
\end{equation}
and conversely,
\begin{equation}
  f_i(\omega) = \int_{-\infty}^\infty d\Omega \left[
    \alpha_{ji}^*(\Omega,\omega) f_j(\Omega) - \beta_{ji}(\Omega,\omega) f_j^*(\Omega)
  \right].
  \label{Eq:3.1.0.2}
\end{equation}
Here, \(\alpha_{ji}\) and \(\beta_{ji}\) are the Bogoliubov coefficients encoding the mixing of positive- and negative-frequency modes between the two bases.

A real scalar quantum field \(\hat{\Phi}\) can be expanded in either basis as
\begin{equation}
 \hat{\Phi} = \int_{-\infty}^\infty d\omega \left[
   \hat{a}_i(\omega) f_i(\omega) + \hat{a}_i^\dagger(\omega) f_i^*(\omega)
 \right],
 \label{Eq:3.1.0.3}
\end{equation}
where the annihilation and creation operators \(\hat{a}_i(\omega)\), \(\hat{a}_i^\dagger(\omega)\) satisfy the equal-time commutation relations
\begin{equation}
    \begin{split}
       & \big[\hat{a}_{i},\hat{a}_{i^{\prime}}\big] =0\\
       & \big[\hat{a}_{i}^\dagger,\hat{a}_{i^\prime}^\dagger\big] =0 \\
        & \big[\hat{a}_{i},\hat{a}_{i^\prime}^\dagger\big] = \delta(i-i^\prime),\label{Eq:3.1.0.4}   
    \end{split}
\end{equation}
valid for free scalar fields quantized on equal-time hypersurfaces in curved spacetime.

The Rindler vacuum $\vert 0_{R_i} \rangle$ associated with the mode set $i$ is defined by
\begin{equation}
    \hat{a}_i(\omega) \vert 0_{R_i} \rangle = 0, \quad \forall \omega.
    \label{Eq:3.1.0.5}
\end{equation}
Similarly, one can expand \(\hat{\Phi}\) in the basis \(\{f_j(\Omega)\}\),
\begin{equation}
 \hat{\Phi} = \int_{-\infty}^\infty d\Omega \left[
   \hat{a}_j(\Omega) f_j(\Omega) + \hat{a}_j^\dagger(\Omega) f_j^*(\Omega)
 \right],
 \label{Eq:3.1.0.6}
\end{equation}
with the annihilation and creation operators $\hat{a}_j(\Omega), \hat{a}_j^\dagger(\Omega)$ satisfying the same commutation relations as in Eq.~(\ref{Eq:3.1.0.4}), and associated vacuum $\vert 0_{R_j}\rangle$ obeying
\begin{equation}
    \hat{a}_j(\Omega) \vert 0_{R_j} \rangle = 0, \quad \forall \Omega.
    \label{Eq:3.1.0.7}
\end{equation}

Using the Bogoliubov transformations (\ref{Eq:3.1.0.1}) and (\ref{Eq:3.1.0.2}), the annihilation and creation operators transform as
\begin{align}
    \hat{a}_j(\Omega) &= \int_0^\infty d\omega \left[
      \alpha_{ji}^*(\Omega,\omega) \hat{a}_i(\omega) - \beta_{ji}^*(\Omega,\omega) \hat{a}_i^\dagger(\omega)
    \right],
    \label{Eq:3.1.0.8} \\
    \hat{a}_j^\dagger(\Omega) &= \int_0^\infty d\omega \left[
      \alpha_{ji}(\Omega,\omega) \hat{a}_i^\dagger(\omega) - \beta_{ji}(\Omega,\omega) \hat{a}_i(\omega)
    \right].
    \label{Eq:3.1.0.9}
\end{align}

When the \(\beta\)-coefficients are non-zero, the particle number operator expectation value evaluated in the vacuum \(\vert 0_{R_i} \rangle\) associated with the \(i\)-modes is
\begin{equation}
    \langle 0_{R_i} \vert \hat{N}_{ji}(\Omega) \vert 0_{R_i} \rangle 
    = \langle 0_{R_i} \vert \hat{a}_j^\dagger(\Omega) \hat{a}_j(\Omega) \vert 0_{R_i} \rangle
    = \int d\omega \, \lvert \beta_{ji}(\Omega,\omega) \rvert^2,
    \label{Eq:3.1.0.10}
\end{equation}
which demonstrates that an observer associated with the \(j\)-modes detects particles in the vacuum defined by the \(i\)-modes whenever \(\beta_{ji} \neq 0\).

In this section, we derive the Bogoliubov transformation between two complete sets of mode functions and use it to evaluate the expectation value of the number operator in different Rindler frames.

\subsection{Virasoro anomaly method}

We consider a massless scalar field in two dimensions. We can therefore use the techniques of CFT in order to evaluate the particle content. We use the standard results involving Virasoro anomaly and the normal-ordered stress tensor~\cite{Fabbri:2005mw}. 
In this work, we employ Rindler coordinates, which cover only a portion of Minkowski spacetime. Nonetheless, the normal-ordered stress tensor is well defined within the Rindler wedge and corresponds to renormalisation with respect to the Rindler vacuum~\cite{Fabbri:2005mw, Blumenhagen:2009zz}. The resulting expectation values have interpretation in terms of observables accessible to uniformly accelerated observers and are directly related to the Unruh effect~\cite{Unruh:1976db,Unruh:1983ms, UtiyamaDeWitt1962}.
Under conformal transformation that maps from a coordinate chart $(U,V)$ to $(u,v)$ via the relation, $U=U(u)$ and similarly $V=V(v)$, the energy-momentum tensor behaves as
\begin{equation*}
    T^\prime_{uu} = \bigg(\frac{\partial U}{\partial u}\bigg)^2 T_{UU} + \frac{c}{12}  S(U,u)
\end{equation*}
Where $S(U,u)$ denotes the Schwarzian derivative and is defined as~\cite{Blumenhagen:2009zz}. And, $c$ denotes the central charge. In the current context, the central charge is $c=-\hbar/2\pi$~\cite{fabbri12005}.
\begin{equation}
    S(U,u) = (1/ \partial_{u} U)^2[(\partial_{u} U)(\partial^3_{u} U)-\frac{3}{2}(\partial^2_{u} U)^2], \label{Eq:4.1.0.3}
\end{equation}
So the relevant equation for the left-moving sector is,
 \begin{equation*}
    T^\prime_{uu} = \bigg(\frac{\partial U}{\partial u}\bigg)^2 T_{UU} - \frac{\hbar}{24\pi}  S(U,u)
\end{equation*}
The corresponding equations for the $V$ sector read,
\begin{equation*}
    T^\prime_{vv} = \bigg(\frac{\partial V}{\partial v}\bigg)^2 T_{VV} - \frac{\hbar}{24\pi}  S(V,v)
\end{equation*} 
where $S(V,v)$ is the Schwarzian derivative defined by replacing the $U,u$s with the $V,v$s. 
\section{Particle content in \texorpdfstring{$R_3$}{R3} along various paths \label{Sec-3} }
    
The explicit calculations and physical interpretations follow in subsequent subsections.

\subsection{Particle content along path 1 \label{Subsec-3.2}}

In this subsection, we evaluate the particle content along Path 1, focusing on the particle content in $R_3$ when the field is in the Minkowski vacuum. The wedge $R_3$ is spatially shifted from the origin by a parameter \(\Delta_1\), as illustrated in Figs.~\ref{Fig:1} and~\ref{Fig:2}. Since the Unruh effect is invariant under such spatial translations, we expect the particle content in $R_3$ to be thermal. We demonstrate this explicitly below.

The light-cone coordinates are related as in Eqs.~(\ref{Eq:2.1.0.7}) and~(\ref{Eq:2.1.0.8}). The right-moving and left-moving components of the Minkowski field operator \(\hat{\phi}_0\) can be expanded as:
\begin{align}
    \hat{\phi}_{0}(U_M) &= \int_0^\infty \frac{d\omega_0}{\sqrt{4\pi \omega_0}} \left[ \overrightarrow{\hat{a}}_0(\omega_0)\, e^{-i \omega_0 U_M} + \overrightarrow{\hat{a}}_0^\dagger(\omega_0)\, e^{i \omega_0 U_M} \right], \label{Eq:3.2.0.1} \\
    \hat{\phi}_{0}(V_M) &= \int_0^\infty \frac{d\omega_0}{\sqrt{4\pi \omega_0}} \left[ \overleftarrow{\hat{b}}_0(\omega_0)\, e^{-i \omega_0 V_M} + \overleftarrow{\hat{b}}_0^\dagger(\omega_0)\, e^{i \omega_0 V_M} \right]. \label{Eq:3.2.0.2}
\end{align}

Starting from Eq.~(\ref{Eq:3.0.0.10}), we multiply both sides by \(\int_{-\infty}^\infty \frac{d u_3}{\sqrt{2\pi}} e^{i \nu' u_3}\), resulting in
 \begin{equation}
  \begin{split}
       \int_{-\infty}^\infty \frac{du_3}{\sqrt{2\pi}} e^{i\;\nu' \;u_{3}}\;\hat{\phi}_3 =  \int_{0}^\infty \frac{d\nu}{\sqrt{2\pi}}\frac{1}{\sqrt{2\nu}} \\
       \int_{-\infty}^\infty \frac{du_3}{\sqrt{2\pi}}\;\bigg(\overrightarrow{\hat{a}}_3(\nu)\; e^{i\;(\nu^\prime-\nu)\; u_{3}}+\overrightarrow{\hat{a}}_3^\dagger(\nu)\; e^{i\;(\nu +\nu^\prime)\;u_{3}}\bigg) , \label{Eq:3.2.0.3}
  \end{split}
  \end{equation}
For $\nu' > 0$, only the first term contributes due to orthogonality, yielding:
\begin{equation}
   \int_{-\infty}^\infty \frac{d u_3}{\sqrt{2\pi}} e^{i \nu' u_3} \hat{\phi}_3 
   = \frac{\overrightarrow{\hat{a}}_3(\nu')}{\sqrt{2 \nu'}}.
   \label{Eq:3.2.0.4}
\end{equation}

We now compute the same integral using the Minkowski field operator $\hat{\phi}_0$ from Eq.~(\ref{Eq:3.2.0.1}),
 \begin{equation}
       \begin{split}
    \int_{-\infty}^\infty \frac{du_3}{\sqrt{2\pi}} e^{i\;\nu'\; u_{3}}\hat{\phi}_{0} =  \frac{1}{2\pi} \int_{0}^\infty \frac{d\omega_0}{\sqrt{2\omega_0}} \int_{-\infty}^\infty du_3\\
    \bigg(\overrightarrow{\hat{a}}_0(\omega_0)\;e^{-i\;\omega_0\; U_M} e^{i\;\nu'\;u_3}+\overrightarrow{\hat{a}_0^\dagger}(\omega_0)\;e^{i\;\omega_0\; U_M}e^{i\;\nu'\;u_3}\bigg),\label{Eq:3.2.0.5}
       \end{split}
   \end{equation}
Setting $\nu' = \nu > 0$, and substituting $U_M$ from Eq.~(\ref{Eq:2.1.0.7}) at $a_3 = a$ and $\Delta_1 = \frac{1}{a}$, we identify the Bogoliubov coefficients by matching Eq.~(\ref{Eq:3.1.0.8}),
\begin{align}
   \overrightarrow{\alpha_{30}^*}(\nu, \omega_0) &= \sqrt{\frac{\nu}{\omega_0}} \int_{-\infty}^\infty \frac{d u_3}{2\pi} e^{\frac{i \omega_0}{a}} \left(e^{-a u_3} + 1\right)^{i \nu / a} e^{i \nu u_3}, \label{Eq:3.2.0.6} \\
   \overrightarrow{\beta_{30}^*}(\nu, \omega_0) &= -\sqrt{\frac{\nu}{\omega_0}} \int_{-\infty}^\infty \frac{d u_3}{2\pi} e^{-\frac{i \omega_0}{a}} \left(e^{-a u_3} + 1\right)^{i \nu / a} e^{i \nu u_3}. \label{Eq:3.2.0.7}
\end{align}

After evaluation, these integrals yield expressions involving Gamma functions:
\begin{align}
 \overrightarrow{\alpha_{30}^*}(\nu, \omega_0) &= \frac{1}{2 \pi a} \sqrt{\frac{\nu}{\omega_0}} e^{\frac{i \omega_0}{a}} \Gamma\left(-\frac{i \nu}{a}\right) \left( \frac{-i \omega_0}{a} \right)^{\frac{i \nu}{a}}, \label{Eq:3.2.0.8} \\
 \overrightarrow{\beta_{30}^*}(\nu, \omega_0) &= -\frac{1}{2 \pi a} \sqrt{\frac{\nu}{\omega_0}} e^{-\frac{i \omega_0}{a}} \Gamma\left(-\frac{i \nu}{a}\right) \left( \frac{i \omega_0}{a} \right)^{\frac{i \nu}{a}}. \label{Eq:3.2.0.9}
\end{align}

The Minkowski vacuum \(\vert 0_M \rangle\) is defined by
\begin{equation}
    \overrightarrow{\hat{a}}_0(\omega_0) \vert 0_M \rangle = 0, \quad \overleftarrow{\hat{b}}_0(\omega_0) \vert 0_M \rangle = 0, \label{Eq:3.2.0.10}
\end{equation}
where the annihilation operators satisfy the commutation relations as per Eq.~(\ref{Eq:3.1.0.4}).

The expectation value of the right-moving particle number operator $\overrightarrow{\hat{N}}_{30}(\nu) = \overrightarrow{\hat{a}}_3^\dagger(\nu) \overrightarrow{\hat{a}}_3(\nu)$ in the Minkowski vacuum is given by
\begin{equation}
    \langle 0_M \vert \overrightarrow{\hat{N}}_{30}(\nu) \vert 0_M \rangle 
    = \int_0^\infty d \omega_0\, \left| \overrightarrow{\beta}_{30}(\nu, \omega_0) \right|^2, \label{Eq:3.2.0.11}
\end{equation}
which simplifies to the Planckian distribution
\begin{equation}
   \langle \overrightarrow{\hat{N}}_{30}(\nu) \rangle = \frac{\delta(0)}{e^{\frac{2 \pi \nu}{a}} - 1}, \label{Eq:3.2.0.12}
\end{equation}
where the formal divergence $\delta(0)$ arises from the infinite volume and is treated as a constant in a finite volume (see Appendix B). The corresponding particle number density reads~\cite{Mukhanov:2007zz}
\begin{equation}
   \overrightarrow{n}_{30}(\nu) = \frac{1}{e^{\frac{2 \pi \nu}{a}} - 1}. \label{Eq:3.2.0.13}
\end{equation}

Thus, the right-moving modes in wedge $R_3$, when evaluated in the Minkowski vacuum, are thermally excited and exhibit a Planck distribution. A similar calculation shows that the left-moving modes also thermalise, confirming full thermality along Path 1.

This thermal mixed state arises because the Minkowski vacuum is entangled between modes inside $R_3$ and those in its complement; tracing over inaccessible modes yields a thermal reduced state.

\subsection{Particle content along path 2}
In this section, we evaluate the particle content along Path 2, which connects the Rindler wedges $R_1$ and $R_3$ through a spatial shift parametrised by $\Delta_1$, as illustrated in Figs.~\ref{Fig:1} and~\ref{Fig:2}. To compute the Bogoliubov coefficients relating the mode functions in these two wedges, we proceed as follows.

Starting from Eq.~(\ref{Eq:3.0.0.10}), we multiply both sides by $\int_{-\infty}^\infty \frac{d u_3}{\sqrt{2\pi}} e^{i \nu' u_3}$, yielding
\begin{equation}
  \begin{split}
       \int_{-\infty}^\infty \frac{du_3}{\sqrt{2\pi}} e^{i\;\nu' \;u_{3}}\;\hat{\phi}_3 =  \int_{0}^\infty \frac{d\nu}{\sqrt{2\pi}}\frac{1}{\sqrt{2\nu}} \\
       \int_{-\infty}^\infty \frac{du_3}{\sqrt{2\pi}}\;\bigg(\overrightarrow{\hat{a}}_3(\nu)\; e^{i\;(\nu^\prime-\nu)\; u_{3}}+\overrightarrow{\hat{a}}_3^\dagger(\nu)\; e^{i\;(\nu +\nu^\prime)\;u_{3}}\bigg) , \label{Eq:3.2.0.15}
  \end{split}
  \end{equation}
where, for $\nu' > 0$, only the first term contributes due to orthogonality, giving
\begin{equation}
\int_{-\infty}^\infty \frac{d u_3}{\sqrt{2\pi}} e^{i \nu' u_3} \hat{\phi}_3 = \frac{\overrightarrow{\hat{a}}_3(\nu')}{\sqrt{2 \nu'}}.
\label{Eq:3.2.0.16}
\end{equation}

Computing the same quantity using the field operator in wedge $R_1$, Eq.~(\ref{Eq:3.0.0.8}), and multiplying by the same factor,
 \begin{equation}
       \begin{split}
    \int_{-\infty}^\infty \frac{du_3}{\sqrt{2\pi}} e^{i\;\nu'\; u_{3}}\hat{\phi}_1 =  \frac{1}{2\pi} \int_{0}^\infty \frac{d\omega}{\sqrt{2\omega}} \int_{-\infty}^\infty du_3\\
    \bigg(\overrightarrow{\hat{a}}_1(\omega)e^{-i\;\omega\; u_1} e^{i\;\nu'\;u_3}+\overrightarrow{\hat{a}}_1^\dagger(\omega)e^{i\;\omega\; u_1}e^{i\;\nu'\;u_3}\bigg),\label{Eq:3.2.0.17}
       \end{split}
   \end{equation}
Setting $\nu' = \nu > 0$ and comparing with Eq.~(\ref{Eq:3.2.0.16}), we obtain:
\begin{equation}
\begin{split}
\overrightarrow{\hat{a}}_3(\nu) = \int_0^\infty d\omega \bigg[
& \overrightarrow{\hat{a}}_1(\omega) \sqrt{\frac{\nu}{\omega}} \int_{-\infty}^\infty \frac{d u_3}{2\pi} e^{-i \omega u_1} e^{i \nu u_3} \\
& + \overrightarrow{\hat{a}}_1^\dagger(\omega) \sqrt{\frac{\nu}{\omega}} \int_{-\infty}^\infty \frac{d u_3}{2\pi} e^{i \omega u_1} e^{i \nu u_3} \bigg].
\end{split}
\label{Eq:3.2.0.18}
\end{equation}

By matching with Eq.~(\ref{Eq:3.1.0.8}), the Bogoliubov coefficients follow as:
\begin{align}
\overrightarrow{\alpha^{*}_{31}}(\nu, \omega) &= \sqrt{\frac{\nu}{\omega}} \int_{-\infty}^\infty \frac{d u_3}{2\pi} e^{-i \omega u_1} e^{i \nu u_3}, \label{Eq:3.2.0.19} \\
\overrightarrow{\beta^{*}_{31}}(\nu, \omega) &= -\sqrt{\frac{\nu}{\omega}} \int_{-\infty}^\infty \frac{d u_3}{2\pi} e^{i \omega u_1} e^{i \nu u_3}. \label{Eq:3.2.0.20}
\end{align}

Using the relation between coordinates $u_1$ and $u_3$ from Eq.~(\ref{Eq:2.1.0.11}) and simplifying, we obtain:
\begin{align}
\overrightarrow{\alpha^{*}_{31}}(\nu, \omega) &= \frac{1}{2\pi a} \sqrt{\frac{\nu}{\omega}} \frac{\Gamma\left(\frac{i\nu}{a}\right) \Gamma\left(-\frac{i(\nu+\omega)}{a}\right)}{\Gamma\left(-\frac{i \omega}{a}\right)}, \label{Eq:3.2.0.21} \\
\overrightarrow{\beta^{*}_{31}}(\nu, \omega) &= -\frac{1}{2\pi a} \sqrt{\frac{\nu}{\omega}} \frac{\Gamma\left(-\frac{i \nu}{a}\right) \Gamma\left(\frac{i(\nu + \omega)}{a}\right)}{\Gamma\left(\frac{i \omega}{a}\right)}. \label{Eq:3.2.0.22}
\end{align}

Squaring the modulus of $\overrightarrow{\beta}_{31}$, we have
\begin{equation}
|\overrightarrow{\beta}_{31}(\nu, \omega)|^2 = \overrightarrow{\beta^{*}_{31}}(\nu, \omega) \overrightarrow{\beta}_{31}(\nu, \omega),
\label{Eq:3.2.0.23}
\end{equation}
which simplifies to
\begin{equation}
|\overrightarrow{\beta}_{31}(\nu, \omega)|^2 = \frac{1}{4 \pi a} \frac{\sinh\left(\frac{\pi \omega}{a}\right)}{(\omega + \nu) \sinh\left(\frac{\pi \nu}{a}\right) \sinh\left(\frac{\pi (\omega + \nu)}{a}\right)}.
\label{Eq:3.2.0.24}
\end{equation}

Defining the vacuum state from Eqs.~(\ref{Eq:3.1.0.5}) and~(\ref{Eq:3.1.0.7}),
\[
\overrightarrow{\hat{a}}_1(\omega) |0_{R_1}\rangle = 0,
\]
where the operators satisfy commutation relations as in Eq.~(\ref{Eq:3.1.0.4}).

The expectation value of the number operator in the vacuum \(|0_{R_1}\rangle\) is~\cite{Padmanabhan:2010xe,Kolekar:2013hra}
\begin{equation}
\langle 0_{R_1} | \overrightarrow{\hat{a}}_3^\dagger(\nu) \overrightarrow{\hat{a}}_3(\nu) | 0_{R_1} \rangle = \int_0^\infty d \omega |\overrightarrow{\beta}_{31}(\nu, \omega)|^2.
\label{Eq:3.2.0.25}
\end{equation}

Substituting Eq.~(\ref{Eq:3.2.0.24}) into the above integral yields
\begin{equation}
\langle \overrightarrow{\hat{N}}_{31}(\nu) \rangle = \frac{1}{4 \pi a \sinh\left(\frac{\pi \nu}{a}\right)} \int_0^\infty d \omega \frac{\sinh\left(\frac{\pi \omega}{a}\right)}{(\omega + \nu) \sinh\left(\frac{\pi (\omega + \nu)}{a}\right)},
\label{Eq:3.2.0.26}
\end{equation}
which, using the integral result from Appendix~\ref{Apn2}, evaluates to
\begin{equation}
\langle \overrightarrow{\hat{N}}_{31}(\nu) \rangle = \frac{\delta(0)}{e^{\frac{2 \pi \nu}{a}} - 1},
\label{Eq:3.2.0.27}
\end{equation}
with \(\delta(0)\) representing the infinite mode density in the unbounded spatial volume~\cite{Frodden:2018mdm}. The corresponding mean particle number density is~\cite{Mukhanov:2007zz}
\begin{equation}
\overrightarrow{n}_{31}(\nu) = \frac{1}{e^{\frac{2 \pi \nu}{a}} - 1}.
\label{Eq:3.2.0.28}
\end{equation}

Hence, the right-moving modes in wedge $R_3$, evaluated from the vacuum of $R_1$, are thermally excited and exhibit a Planckian distribution. A similar calculation for the left-moving modes confirms their thermal excitation, establishing full thermalisation of both modes along Path 2.

The resulting mixed state arises because the vacuum $|0_{R_1}\rangle$ is entangled between modes inside the wedge and their complementary modes, with tracing over inaccessible regions yielding a thermal reduced state in $R_3$.

\subsection{ Path 3. Particle content along shifts along \texorpdfstring{$V_1$}{V1} and  \texorpdfstring{$U_2$}{U2} axes respectively \label{Subsec-3.3}}

In this section, we evaluate the Bogoliubov transformations and the resulting particle content along Path 3, where the Rindler wedge $\mathrm{R}_2$ is obtained from $\mathrm{R}_1$ by a null shift along the $V_1$-axis, and $\mathrm{R}_3$ is subsequently obtained from $\mathrm{R}_2$ along the $U_2$-axis, as illustrated in Fig.~\ref{Fig:1}. Thus, the evaluation of particle content in $R_3$ from the vacuum in $R_1$ proceeds in two stages, with the intermediate wedge $R_2$ serving as a pitstop.

To calculate the required Bogoliubov coefficients, we begin by multiplying both sides of Eq.~(\ref{Eq:3.0.0.7}) by the Fourier mode \[\int_{-\infty}^\infty \frac{du_2}{\sqrt{2\pi}} e^{i \Omega' u_2}.\] After simplifications and using the orthogonality of exponentials, we obtain~\cite{Padmanabhan:2010zzb}
 \begin{equation}
  \begin{split}
       \int_{-\infty}^\infty \frac{du_2}{\sqrt{2\pi}}\; e^{i\;\Omega^\prime\; u_{2}}\;\hat{\phi}_2 &=  \int_{0}^\infty \frac{d\Omega}{\sqrt{2\pi}}\frac{1}{\sqrt{2\Omega}} \\
       \int_{-\infty}^\infty \frac{du_2}{\sqrt{2\pi}}\;\bigg(\overrightarrow{\hat{a}}_2(\Omega)\; e^{i\;(\Omega^\prime-\Omega)\; u_{2}}&+\overrightarrow{\hat{a}}_2^\dagger(\Omega) \;e^{i\;(\Omega +\Omega^\prime)\;u_{2}}\bigg). \label{Eq:3.3.0.1}
  \end{split}
  \end{equation}
Which reduces to
\begin{equation}
\int_{-\infty}^\infty \frac{du_2}{\sqrt{2\pi}} e^{i \Omega' u_2} \hat{\phi}_2 = \frac{\overrightarrow{\hat{a}}_2(\Omega')}{\sqrt{2\Omega'}},
\label{Eq:3.3.0.2}
\end{equation}
due to orthogonality, valid for $\Omega' > 0$.
Next, we compute the same integral using the field operator $\hat{\phi}_1(u_1)$, expressed in terms of $u_2$ via the null-shift relation connecting wedges $R_1$ and $R_2$. Multiplying both sides of Eq.~(\ref{Eq:3.0.0.6}) by the same Fourier kernel gives
 \begin{equation}
       \begin{split}
     \int_{-\infty}^\infty \frac{du_2}{\sqrt{2\pi}}\; e^{i\;\Omega^\prime\; u_{2}}\;\hat{\phi}_1 =   \frac{1}{2\pi} \int_{0}^\infty \frac{d\omega}{\sqrt{2\omega}}
     \bigg(\overrightarrow{\hat{a}}_1(\omega)\int_{-\infty}^\infty du_2\\ \;e^{i\;\Omega^\prime\; u_{2}-i\;\omega\; u_{1}} +\overrightarrow{\hat{a}}_1^\dagger(\omega)\;\int_{-\infty}^\infty du_2 \;e^{i\;\Omega^\prime\; u_{2}+i\;\omega\; u_{1}} \bigg).\label{Eq:3.3.0.3}     
       \end{split}
   \end{equation}
Matching Eqs.~(\ref{Eq:3.3.0.2}) and (\ref{Eq:3.3.0.3}), and evaluating at $\Omega' = \Omega$, allows us to identify the Bogoliubov coefficients between wedges $\mathrm{R}_1$ and $\mathrm{R}_2$:
\begin{align}
\overrightarrow{\alpha^{*}_{21}}(\Omega, \omega) &= \sqrt{\frac{\Omega}{\omega}} \int_{-\infty}^\infty \frac{du_2}{2\pi} e^{i \Omega u_2 - i \omega u_1}, \label{Eq:3.3.0.5} \\
\overrightarrow{\beta^{*}_{21}}(\Omega, \omega) &= -\sqrt{\frac{\Omega}{\omega}} \int_{-\infty}^\infty \frac{du_2}{2\pi} e^{i \Omega u_2 + i \omega u_1}. \label{Eq:3.3.0.6}
\end{align}
Utilizing the explicit coordinate transformation between $u_1$ and $u_2$ from Eq.~(\ref{Eq:2.1.1.9}), one finds that
\begin{equation}
\overrightarrow{\alpha^{*}_{21}}(\Omega, \omega) = \delta(\Omega - \omega) \sqrt{\frac{\Omega}{\omega}},
\label{Eq:3.3.0.7}
\end{equation}
and
\begin{equation}
\overrightarrow{\beta^{*}_{21}}(\Omega, \omega) = 0.
\label{Eq:3.3.0.8}
\end{equation}
This confirms that, for the right-moving modes, the null shift from $\mathrm{R}_1$ to $\mathrm{R}_2$ does not excite particles; the vacuum structure remains preserved.

We now compute the Bogoliubov coefficients for the left-moving modes relating the Rindler wedges $\mathrm{R}_1$ and $\mathrm{R}_2$. Starting from the transformation relations and Eq.~(\ref{Eq:3.1.0.8}), the coefficient $\overleftarrow{\alpha^{*}_{21}}(\Omega,\omega)$ is expressed as
\begin{equation}
  \overleftarrow{\alpha^{*}_{21}}(\Omega,\omega) = \frac{1}{2\pi} \sqrt{\frac{\Omega}{\omega}} \int_{-\infty}^\infty dv_2 \, e^{i \Omega v_2} \, e^{-\frac{i \omega}{a} \ln\left[e^{a v_2} + 1\right]},
  \label{Eq:3.3.0.9}
\end{equation}
which simplifies through standard integration techniques (see Appendix A) to
\begin{equation}
 \overleftarrow{\alpha^{*}_{21}}(\Omega,\omega) = \frac{1}{2\pi a} \sqrt{\frac{\Omega}{\omega}} \frac{2^{\frac{-i\omega + i\Omega}{a}} \; \Gamma\left(\frac{i \Omega}{a}\right) \; \Gamma\left(\frac{i \omega - i \Omega}{a}\right)}{\Gamma\left(\frac{i \omega}{a}\right)}.
 \label{Eq:3.3.0.10}
\end{equation}
Similarly, the Bogoliubov coefficient $\overleftarrow{\beta^{*}_{21}}(\Omega,\omega)$ is given by
\begin{equation}
  \overleftarrow{\beta^{*}_{21}}(\Omega,\omega) = -\frac{1}{2\pi} \sqrt{\frac{\Omega}{\omega}} \int_{-\infty}^\infty dv_2 \, e^{i \Omega v_2} \, e^{i \omega v_1},
  \label{Eq:3.3.0.11}
\end{equation}
and evaluates to
\begin{equation}
  \overleftarrow{\beta^{*}_{21}}(\Omega,\omega) = -\frac{1}{2\pi a} \sqrt{\frac{\Omega}{\omega}} \frac{2^{\frac{i(\omega + \Omega)}{a}} \; \Gamma\left(\frac{i \Omega}{a}\right) \; \Gamma\left(-\frac{i(\omega + \Omega)}{a}\right)}{\Gamma\left(-\frac{i \omega}{a}\right)}.
  \label{Eq:3.3.0.12}
\end{equation}
The squared modulus of the Bogoliubov coefficient is
\begin{equation}
  \left| \overleftarrow{\beta_{21}}(\Omega,\omega) \right|^{2} = \overleftarrow{\beta^{*}_{21}}(\Omega,\omega) \; \overleftarrow{\beta_{21}}(\Omega,\omega),
  \label{Eq:3.3.0.13}
\end{equation}
Which simplifies to
\begin{equation}
  \left| \overleftarrow{\beta_{21}}(\Omega,\omega) \right|^{2} = \frac{1}{4 \pi^{2} a^{2}} \frac{\Omega}{\omega} \frac{\left| \Gamma\left(\frac{i \Omega}{a}\right) \right|^{2} \left| \Gamma\left(\frac{i (\Omega + \omega)}{a}\right) \right|^{2}}{\left| \Gamma\left(\frac{i \omega}{a}\right) \right|^{2}}.
  \label{Eq:3.3.0.14}
\end{equation}

Applying standard identities for Gamma functions, the above expression reduces to
\begin{equation}
  \left| \overleftarrow{\beta_{21}}(\Omega,\omega) \right|^{2} = \frac{1}{4 \pi a} \frac{\sinh\left( \frac{\pi \omega}{a} \right)}{(\omega + \Omega) \sinh\left( \frac{\pi \Omega}{a} \right) \sinh\left[ \frac{\pi (\omega + \Omega)}{a} \right]}.
  \label{Eq:3.3.0.15}
\end{equation}

The Rindler vacuum in wedge $\mathrm{R}_1$ is defined by the annihilation condition $\overleftarrow{\hat{b}}_1 |0_{R_1} \rangle = 0,$
and factorises as
$
|0_{R_1}\rangle = |\overleftarrow{0}_{R_1}\rangle \otimes |\overrightarrow{0}_{R_1}\rangle,
$
where $|\overleftarrow{0}_{R_1}\rangle$ and $|\overrightarrow{0}_{R_1}\rangle$ denote the left- and right-moving vacuum states, respectively. The operators $\overleftarrow{\hat{b}}_1$ and $\overleftarrow{\hat{b}}_1^\dagger$ satisfy canonical commutation relations analogous to those in Eq.~(\ref{Eq:3.1.0.4}).

We then compute the particle content in the left-moving modes of $\mathrm{R}_2$ when the field is in the $\mathrm{R}_1$ vacuum. The expectation value of the number operator is
\begin{equation}
  \langle 0_{R_1} | \overleftarrow{\hat{b}}_2^\dagger(\Omega) \overleftarrow{\hat{b}}_2(\Omega) | 0_{R_1} \rangle = \int_0^\infty d\omega \left| \overleftarrow{\beta_{21}}(\Omega, \omega) \right|^{2},
  \label{Eq:3.3.0.16}
\end{equation}
Which, upon substituting Eq.~(\ref{Eq:3.3.0.15}), becomes
\begin{equation}
  \langle \overleftarrow{\hat{N}}_{21}(\Omega) \rangle = \frac{1}{4 \pi a \sinh\left( \frac{\pi \Omega}{a} \right)} \int_0^\infty d\omega \frac{\sinh\left( \frac{\pi \omega}{a} \right)}{(\omega + \Omega) \sinh\left[ \frac{\pi (\omega + \Omega)}{a} \right]}.
  \label{Eq:3.3.0.17}
\end{equation}
This integral is analytically tractable (see Appendix.~\ref{Apn2}) and evaluates to
\begin{equation}
  \langle \overleftarrow{\hat{N}}_{21}(\Omega) \rangle = \frac{\delta(0)}{e^{\frac{2 \pi \Omega}{a}} - 1}.
  \label{Eq:3.3.0.18}
\end{equation}

Here, $\delta(0)$ formally arises from the infinite volume limit and can be interpreted as a constant density over a finite spacetime volume~\cite{Frodden:2018mdm}. Accordingly, the mean particle number density for frequency $\Omega$ is given by the Planckian distribution~\cite{Mukhanov:2007zz}:
\begin{equation}
  \overleftarrow{n}_{21}(\Omega) = \frac{1}{e^{\frac{2 \pi \Omega}{a}} - 1}.
  \label{Eq:3.3.0.19}
\end{equation}

This confirms that the left-moving modes are thermally excited during the transition from $\mathrm{R}_1$ to $\mathrm{R}_2$, at a characteristic Unruh-like temperature $T = \frac{a}{2 \pi}$. In contrast, the right-moving modes remain in the vacuum state.
We therefore have the left-moving modes in a thermal state, and the right-moving modes have no excitations. This implies that the right-moving modes are in the ground state for all momenta $k$,
\begin{equation}
    |\overrightarrow{0_k \rangle}_{R_1}=|\overrightarrow{0_k \rangle}_{R_2}
\end{equation}

We now evaluate the particle content of the right-moving modes in the transition from Rindler wedge $\mathrm{R}_2$ to wedge $\mathrm{R}_3$. The mode functions in $\mathrm{R}_3$ are related to those in $\mathrm{R}_2$. Setting \(j=3\), \(i=2\), and identifying \(\omega = \Omega\), \(\nu = \Omega\) in Eqs.~(\ref{Eq:3.1.0.8}) and ~(\ref{Eq:3.1.0.9}), the Bogoliubov coefficients for the right-moving mode are given by:
\begin{align}
\overrightarrow{\alpha^{*}_{32}}(\nu, \Omega) &= \sqrt{\frac{\nu}{\Omega}} \int_{-\infty}^\infty \frac{d u_3}{2\pi} e^{i \nu u_3} e^{-i \Omega u_2}, \label{Eq:3.3.0.20} \\
\overrightarrow{\beta^{*}_{32}}(\nu, \Omega) &= -\sqrt{\frac{\nu}{\Omega}} \int_{-\infty}^\infty \frac{d u_3}{2\pi} e^{i \nu u_3} e^{i \Omega u_2}. \label{Eq:3.3.0.21}
\end{align}
Using the coordinate transformation from Eq.~(\ref{Eq:2.1.1.19}), Eq.~(\ref{Eq:3.3.0.20}) becomes:
\begin{equation}
\overrightarrow{\alpha^{*}_{32}}(\nu, \Omega) = \frac{1}{2\pi} \sqrt{\frac{\nu}{\Omega}} \int_{-\infty}^\infty d u_3 \, e^{i \nu u_3} \, e^{\frac{i \Omega}{a} \ln\left( e^{-a u_3} + 1 \right)},
\label{Eq:3.3.0.22}
\end{equation}
which evaluates, using standard techniques (see Appendix~\ref {Apn1}), to
\begin{equation}
\overrightarrow{\alpha^{*}_{32}}(\nu, \Omega) = \frac{1}{2 \pi a} \sqrt{\frac{\nu}{\Omega}} \frac{2^{\frac{i(\Omega - \nu)}{a}} \Gamma\left( -\frac{i \nu}{a} \right) \Gamma\left( \frac{i(\nu - \Omega)}{a} \right)}{\Gamma\left( -\frac{i \Omega}{a} \right)}.
\label{Eq:3.3.0.23}
\end{equation}
Similarly, using Eq.~(\ref{Eq:2.1.1.22}), Eq.~(\ref{Eq:3.3.0.21}) becomes:
\begin{equation}
\overrightarrow{\beta^{*}_{32}}(\nu, \Omega) = - \frac{1}{2 \pi} \sqrt{\frac{\nu}{\Omega}} \int_{-\infty}^\infty d u_3 \, e^{i \nu u_3} \, e^{-\frac{i \Omega}{a} \ln\left( e^{-a u_3} + 1 \right)},
\label{Eq:3.3.0.24}
\end{equation}
which evaluates to
\begin{equation}
\overrightarrow{\beta^{*}_{32}}(\nu, \Omega) = - \frac{1}{2 \pi a} \sqrt{\frac{\nu}{\Omega}} \frac{2^{-\frac{i(\Omega - \nu)}{a}} \Gamma\left( -\frac{i \nu}{a} \right) \Gamma\left( \frac{i (\nu + \Omega)}{a} \right)}{\Gamma\left( \frac{i \Omega}{a} \right)}.
\label{Eq:3.3.0.25}
\end{equation}
The squared modulus of the Bogoliubov coefficient is
\begin{equation}
\left| \overrightarrow{\beta_{32}}(\nu, \Omega) \right|^2 = \overrightarrow{\beta^{*}_{32}}(\nu, \Omega) \, \overrightarrow{\beta_{32}}(\nu, \Omega),
\label{Eq:3.3.0.26}
\end{equation}
which simplifies to
\begin{equation}
\left| \overrightarrow{\beta_{32}}(\nu, \Omega) \right|^2 = \frac{1}{4 \pi a} \frac{\sinh\left( \frac{\pi \Omega}{a} \right)}{(\nu + \Omega) \sinh\left( \frac{\pi \nu}{a} \right) \sinh\left[ \frac{\pi (\nu + \Omega)}{a} \right]}.
\label{Eq:3.3.0.28}
\end{equation}
Since \(\mathrm{R}_2\) is obtained from \(\mathrm{R}_1\) via a null shift, the resulting quantum field configuration exhibits selective thermalisation: the left-moving modes become thermally populated, while the right-moving modes remain in the vacuum, as previously established.
We now compute the expectation value of the number operator for right-moving modes in \(\mathrm{R}_3\) evaluated in the state \(\mid \overrightarrow{0}_{R_2} \rangle\). The number operator
\[
\overrightarrow{\hat{N}}_{32}(\nu) = \overrightarrow{\hat{a}}_3^\dagger(\nu) \overrightarrow{\hat{a}}_3(\nu)
\]
has expectation value
\begin{equation}
\langle \overrightarrow{\hat{N}}_{32}(\nu) \rangle = \langle \overrightarrow{0}_{R_2} \mid \overrightarrow{\hat{a}}_3^\dagger(\nu) \overrightarrow{\hat{a}}_3(\nu) \mid \overrightarrow{0}_{R_2} \rangle.
\label{Eq:3.3.0.30}
\end{equation}
Expressed in terms of Bogoliubov coefficients, this is
\begin{equation}
\begin{split}
\langle \overrightarrow{\hat{N}}_{32}(\nu) \rangle = \int_0^\infty d\Omega \int_0^\infty d\Omega' \, \overrightarrow{\beta^{*}_{32}}(\nu, \Omega) \overrightarrow{\beta}_{32}(\nu, \Omega') \\
\times \langle \overrightarrow{0}_{R_2} \mid \overrightarrow{\hat{a}}_2(\Omega') \overrightarrow{\hat{a}}_2^\dagger(\Omega) \mid \overrightarrow{0}_{R_2} \rangle.
\end{split}
\label{Eq:3.3.0.31}
\end{equation}
This implies,
\begin{equation}
\begin{split}
    \langle \overrightarrow{0}_{R_2} \mid \overrightarrow{\hat{a}}_2(\Omega') \overrightarrow{\hat{a}}_2^\dagger(\Omega) \mid \overrightarrow{0}_{R_2} \rangle = \langle \overrightarrow{0} \mid \overrightarrow{\hat{a}}_2(\Omega') \overrightarrow{\hat{a}}_2^\dagger(\Omega) \mid \overrightarrow{0} \rangle \\= \delta(\Omega - \Omega').
\end{split}
\label{Eq:3.3.0.32}
\end{equation}
Therefore, the expectation value reduces to
\begin{equation}
\langle \overrightarrow{\hat{N}}_{32}(\nu) \rangle = \int_0^\infty d\Omega \, \left| \overrightarrow{\beta_{32}}(\nu, \Omega) \right|^2,
\label{Eq:3.3.0.34}
\end{equation}
which, by substitution of Eq.~(\ref{Eq:3.3.0.28}), is an integral of the same form as in Eq.~(\ref{Eq:3.3.0.17}). Evaluating this integral (see Appendix B) gives
\begin{equation}
\langle \overrightarrow{\hat{N}}_{32}(\nu) \rangle = \frac{\delta(0)}{e^{\frac{2 \pi \nu}{a}} - 1}.
\label{Eq:3.3.0.35}
\end{equation}
The factor \(\delta(0)\) arises from the mode normalisation in the continuum and represents an infinite constant related to the volume. Hence, the mean particle number density per unit frequency \(\nu\) in \(\mathrm{R}_3\), as measured in \(\mathrm{R}_2\), is
\begin{equation}
\overrightarrow{n}_{32}(\nu) = \frac{1}{e^{\frac{2 \pi \nu}{a}} - 1},
\label{Eq:3.3.0.36}
\end{equation}
expressing the familiar Planckian distribution at the Unruh like temperature $T = \frac{a}{2 \pi}$.
We conclude that the right-moving modes mediating between $\mathrm{R}_2$ and $\mathrm{R}_3$ exhibit a thermal population characterised by this Planckian spectrum when the field in $\mathrm{R}_2$ is selectively excited.

We now evaluate the particle content of the left-moving modes in the transition from Rindler wedge  $\mathrm{R}_2$ to wedge $\mathrm{R}_3$. The Bogoliubov coefficients relating the mode functions across these regions are given by:
\begin{equation}
  \overleftarrow{\alpha^{*}_{32}}(\nu,\Omega)= \sqrt{\frac{\nu}{\Omega}}\int_{-\infty}^\infty \frac{dv_3}{2\pi}\;e^{i\nu v_{3}}\; e^{-i\Omega v_{2}},\label{Eq:3.3.0.37}
  \end{equation} 
  
  \begin{equation}
     \overleftarrow{\beta^{*}_{32}}(\nu,\Omega)= -\sqrt{\frac{\nu}{\Omega}}\int_{-\infty}^\infty \frac{dv_3}{2\pi}\;e^{i\nu v_{3}}\; e^{i\Omega v_{2}},\label{Eq:3.3.0.38}
  \end{equation}
  The coordinate relation between the wedges, as given by  Eq.~(\ref{Eq:2.1.1.20}), implies $v_2=v_3$.Substituting this into the expressions above and evaluating the integrals, we find:
  \begin{equation}
  \overleftarrow{\alpha^{*}_{32}}(\nu,\Omega)= \sqrt{\frac{\nu}{\Omega}}\;\delta(\nu-\Omega),\label{Eq:3.3.0.39}
  \end{equation} 
  and
  \begin{equation}
    \overleftarrow{\beta^{*}_{32}}(\nu,\Omega)=  0 ,\label{Eq:3.3.0.40}
  \end{equation}
  Hence, the left-moving modes in wedge $\mathrm{R}_2$  remain in vacuum when mapped to a wedge $\mathrm{R}_3$;  no particle creation occurs in this region. This is consistent with the preservation of the positive-frequency structure of the modes under the coordinate transformation. So both the left-moving and right-moving modes have a thermal distribution of particles

 \subsection{ Path 4: Particle content along the shifts along   \texorpdfstring{$U_1$}{U1} and  \texorpdfstring{$V_2$}{V2} axes\label{Subsec-3.4}}
In this section, we analyze the Bogoliubov transformations and particle content along Path 4, wherein the Rindler wedge $\mathrm{R}_4$ is obtained from $\mathrm{R}_1$ via a null shift along the $U_1$-axis, followed by the transition $\mathrm{R}_4 \to \mathrm{R}_3$ through a null shift along the $V_2$-axis, as depicted in Fig.~\ref{Fig:2}.
Since the method for computing Bogoliubov coefficients has already been established in the previous section, we directly write down the expressions for the transition $\mathrm{R}_1 \to \mathrm{R}_4$, which follows the same computational steps. The Bogoliubov coefficients for the right-moving modes are given by:
\begin{equation}
  \overrightarrow{\alpha^{*}_{41}}(\rho,\omega)= \sqrt{\frac{\rho}{\omega}}\int_{-\infty}^\infty \frac{du_4}{2\pi}\; e^{i\rho u_{4}}\; e^{-i\omega u_{1}},\label{Eq:3.4.1.1}
  \end{equation} 
  \begin{equation}
    \overrightarrow{\beta^{*}_{41}}(\rho,\omega)= -\sqrt{\frac{\rho}{\omega}}\int_{-\infty}^\infty \frac{du_4}{2\pi}\;e^{i\rho u_{4}} \;e^{i\omega u_{1}},\label{Eq:3.4.1.2}
  \end{equation} 
  Where the coordinates,$ u_1$ and $u_4$ are related via Eq.~(\ref{Eq:2.1.1.22}).
 As these integrals follow the same structure as in the previous section, we state the result directly. The mean number of particles in the right-moving mode of $\mathrm{R}_4$, when the field is in the vacuum associated with $\mathrm{R}_1$,is given by:
\begin{equation}
   \overrightarrow{n}_{41}(\rho) =  \frac{1}{e^{\frac{2\;\pi\;\rho}{a}}-1}, \label{Eq:3.4.1.3}
\end{equation}
Which is a Planckian distribution at temperature $T=\frac{a}{2\pi}$ arises due to a horizon effect induced by null shifts. This can be regarded as a certain converse of the Unruh effect.

We now evaluate the Bogoliubov coefficients for the left-moving modes associated with the transition $ R_1 \to R_4$:
 \begin{equation}
  \overleftarrow{\alpha^{*}_{41}}(\rho,\omega)= \sqrt{\frac{\rho}{\omega}}\int_{-\infty}^\infty \frac{dv_4}{2\pi}\;e^{i\Omega v_{4}}\; e^{-i\omega v_{1}},\label{Eq:3.4.1.4}
  \end{equation} 
  \begin{equation}
    \overleftarrow{\beta^{*}_{41}}(\rho,\omega)= -\sqrt{\frac{\rho}{\omega}}\int_{-\infty}^\infty \frac{dv_4}{2\pi}\;e^{i\rho v_{4}} \;e^{i\omega v_{1}},\label{Eq:3.4.1.5}
  \end{equation} 
  Substituting $ v_1 = v_4$ from Eq.~(\ref{Eq:2.1.1.22}), the integrals simplify to:
   \begin{equation}
      \overleftarrow{\alpha^{*}_{41}}(\rho,\omega) = \delta(\rho-\omega)\;\sqrt{\frac{\rho}{\omega}},  \label{Eq:3.4.1.6}
  \end{equation}
\begin{equation}
      \overleftarrow{\beta^{*}_{41}}(\rho,\omega) = 0,  \label{Eq:3.4.1.7}
  \end{equation}
This confirms that no particle creation occurs in the left-moving sector during the transition from $ R_1 \to R_4 $.
We conclude that the null shift along the $ U_1$-axis selectively excites only the right-moving modes in $\mathrm{R}_4$, while the left-moving modes remain in the vacuum. We therefore have,
\begin{equation}
    |\overleftarrow{0_k\rangle}_{R_1}=|\overleftarrow{0_k\rangle}_{R_4},
\end{equation}
 
We now evaluate the  Bogoliubov transformation from $\mathrm{R}_4$ to $\mathrm{R}_3$ for the right-moving mode between regions.  As before, the coefficients take the form:
  \begin{equation}
  \overrightarrow{\alpha^{*}_{34}}(\nu,\rho)= \sqrt{\frac{\nu}{\rho}}\int_{-\infty}^\infty \frac{du_3}{2\pi}\;e^{i\nu u_{3}}\; e^{-i\rho u_{4}},\label{Eq:3.4.1.8}
  \end{equation} 
  \begin{equation}
     \overrightarrow{\beta^{*}_{34}}(\nu,\rho)= -\sqrt{\frac{\nu}{\rho}}\int_{-\infty}^\infty \frac{du_3}{2\pi}\;e^{i\nu u_{3}}\; e^{i\rho u_{4}},\label{Eq:3.4.1.9}
  \end{equation} 
   Using the coordinate transformation from Eq.~(\ref{Eq:2.1.1.27}) that relates $u_3$ and $u_4$,we find that: 
   \begin{equation}
     \overrightarrow{\alpha^{*}_{34}}(\nu,\rho) = \delta(\nu-\rho)\;\sqrt{\frac{\nu}{\rho}},  \label{Eq:3.4.1.10}
  \end{equation}
\begin{equation}
     \overrightarrow{\beta^{*}_{34}}(\nu,\rho) = 0.  \label{Eq:3.4.1.11}
  \end{equation}
Hence, the right-moving modes in wedge $R_4$ remain in vacuum when mapped to wedge $R_3$; no particle creation occurs in this region.

We now evaluate the Bogoliubov coefficients for the left-moving mode associated with the transition from $\mathrm{R}_4$ to $\mathrm{R}_3$.The mode functions are related via  Eqs.~(\ref{Eq:3.1.0.8}) and ~(\ref{Eq:3.1.0.9}),with $j = 3$, $i = 2$, $\Omega = \nu$, and $\omega = \rho$. The Bogoliubov coefficients take the form:
  \begin{equation}
  \overleftarrow{\alpha^{*}_{34}}(\nu,\rho)= \sqrt{\frac{\nu}{\rho}}\int_{-\infty}^\infty \frac{dv_3}{2\pi}\;e^{i\nu v_{3}}\; e^{-i\rho v_{4}},\label{Eq:3.4.1.12}
\end{equation} 
\begin{equation}
    \overleftarrow{\beta^{*}_{34}}(\nu,\rho)= -\sqrt{\frac{\nu}{\rho}}\int_{-\infty}^\infty \frac{dv_3}{2\pi}\;e^{i\nu v_{3}} \;e^{i\rho v_{4}}.\label{Eq:3.4.1.13}
\end{equation} 
Substituting the coordinate transformation from  Eq.~(\ref{Eq:2.1.1.28})into the integrals and performing the simplification, we obtain: 
\begin{equation}
   \overleftarrow{\alpha^{*}_{34}}(\nu,\Omega)=\frac{1}{2\pi a} \sqrt{\frac{\nu}{\rho}}\frac{2^{\frac{-i(\rho-\nu)}{a}}\;\Gamma[\frac{i\nu}{a}]\;\Gamma[\frac{i(\rho-\nu)}{a}]}{\Gamma[\frac{i\rho}{a}]},\label{Eq:3.4.1.14}  
\end{equation}  
\begin{equation}
\overleftarrow{\beta^{*}_{34}}(\nu,\rho)=-\frac{1}{2\pi a} \sqrt{\frac{\nu}{\rho}}\frac{2^{\frac{i(\rho+\nu)}{a}}\;\Gamma[\frac{i\nu}{a}]\;\Gamma[\frac{-i(\rho+\nu)}{a}]}{\Gamma[\frac{-i\rho}{a}]}.\label{Eq:3.4.1.15}  
\end{equation}
From the definition:
\begin{equation}
    {\mid \overleftarrow{\beta_{34}}(\nu,\rho) \mid}^{2} = \overleftarrow{\beta^{*}_{34}}(\nu,\rho)\;\overleftarrow{\beta_{34}}(\nu,\rho).\label{Eq:3.4.1.16}
\end{equation}
 Substituting Eq.~(\ref{Eq:3.4.1.15}) into  Eq.~(\ref{Eq:3.4.1.16}),we obtain:
\begin{equation}
  {\mid \overleftarrow{\beta_{34}}(\nu,\rho) \mid}^{2} = \frac{1}{4\pi^2 a^2} \;\frac{\nu}{\rho}\; \frac{{\mid \Gamma[\frac{i\nu}{a}] \mid}^{2}\;{\mid \Gamma[\frac{i(\nu+\rho)}{a}] \mid}^{2}}{ {\mid \Gamma[\frac{i\rho}{a}] \mid}^{2}}. \label{Eq:3.4.1.17}
\end{equation}
Simplifying the above expression using properties of the Gamma function yields:
\begin{equation}
 {\mid\overleftarrow{\beta_{34}}(\nu,\rho} )\mid^{2} =  \frac{1}{4\pi a}\frac{sinh(\frac{\pi\rho}{a})}{(\nu+\rho)\;sinh(\frac{\pi\nu}{a})\;sinh[\frac{\pi(\nu+\rho)}{a}]}.\label{Eq:3.4.1.18}   
\end{equation}
Since $\mathrm{R}_4$ is obtained through a null shift from $\mathrm{R}_1$, the resulting quantum field configuration exhibits selective thermalisation, and the right-moving modes become thermally populated. In contrast, the left-moving modes remain in the vacuum, as established earlier. 

We now compute the expectation value of the number operator for left-moving modes in the region $\mathrm{R}_3$, to the quantum state in region  $\mathrm{R}_4$. 
Substituting the Bogoliubov transformation relations from  Eqs.~(\ref{Eq:3.1.0.8}) and  ~(\ref{Eq:3.1.0.9}) into  Eq.~(\ref{Eq:3.4.1.16}) and using the (j=3 and i=4 and also $\Omega=\nu$ and $\omega=\rho$ ),we obtain:
\begin{equation}
  \begin{split}
     \langle\overleftarrow{\hat{N}}_{34}(\nu)\rangle = \langle  \overrightarrow{0}_{R_4}\mid\int_{0}^{\infty} d\rho \int_{0}^{\infty} d\rho^\prime\;\overleftarrow{\beta^{*}_{34}}(\nu,\rho)\;\overleftarrow{\beta_{34}}(\nu,\rho^\prime)\\
     \overleftarrow{\hat{b}}_{4}(\rho^\prime)\;\overleftarrow{\hat{b}}_{4}^\dagger(\rho) \mid \overrightarrow{0}_{R_4}\rangle, \label{Eq:3.4.1.21}
  \end{split}
\end{equation}
The vacuum expectation value becomes:
\begin{equation}
  \langle  \overrightarrow{0}_{R_4}\mid \;\overleftarrow{\hat{b}}_{4}(\rho^\prime)\;\overleftarrow{\hat{b}}_{4}^\dagger(\rho) \mid   \overrightarrow{0}_{R_4}\rangle  = {\langle \overleftarrow{0}} \mid\overleftarrow{\hat{b}}_{4}(\rho^\prime)\;\overleftarrow{\hat{b}}_{4}^\dagger(\rho) \mid\overleftarrow{0}\rangle ,  \label{Eq:3.4.1.22}
\end{equation}
Since the left-moving modes are in vacuum, we obtain:
\begin{equation}
 {\langle \overleftarrow{0}} \mid \overleftarrow{\hat{b}}_{4}(\rho^\prime) \;\overleftarrow{\hat{b}}_{4}^\dagger(\rho)\mid\overleftarrow{0}\rangle =\delta(\rho-\rho^\prime). \label{Eq:3.4.1.23} 
\end{equation}
From Eqs.~(\ref{Eq:3.4.1.21})–(\ref{Eq:3.4.1.23}), we obtain
\begin{equation}
     \langle \overleftarrow{\hat{N}}_{34}(\nu)\rangle =\int_{0}^{\infty} d\rho \;{\mid\overleftarrow{\beta_{34}}(\nu,\rho)\mid}^2, \label{Eq:3.4.1.24}
\end{equation}
Substituting Eq.~(\ref{Eq:3.4.1.18}) into Eq.~(\ref{Eq:3.4.1.24}) yields an expression structurally identical to Eq.~(\ref{Eq:3.2.0.15}), and we obtain:
\begin{equation}
\langle \overleftarrow{\hat{N}}_{34}(\nu)\rangle=  \frac{\delta(0)}{e^{\frac{2\;\pi\;\nu}{a}}-1},\label{Eq:3.4.1.25}
\end{equation}
where the delta function $\delta(0)$ arises due to the normalisation of the continuous frequency modes and indicates a formally infinite total density per unit frequency. We can therefore identify the mean particle number density per unit frequency $\nu$ in the $\mathrm{R}_3$ vacuum as observed in $\mathrm{R}_4$ as:
\begin{equation}
   \overleftarrow{n}_{34}(\nu) = \frac{1}{e^{\frac{2\;\pi\;\nu}{a}}-1} \label{Eq:3.4.1.26},
\end{equation}
This corresponds to a Planckian spectrum at a temperature of $T = \frac{a}{2\pi}$. Thus, we conclude that the left-moving modes connecting regions $\mathrm{R}_4$ and $\mathrm{R}_3$ are thermally populated with a Planckian distribution, as perceived by observers in $\mathrm{R}_3$.

\section{Particle content via Virasoro anomaly\label{Sec-4}}  
 In this section, we investigate the thermal properties of the nested Rindler wedges shown in Figs.~\ref{Fig:1} and ~\ref{Fig:2} through the expectation values of the energy–momentum tensor in a two-dimensional conformal field theory. Our approach begins with the Minkowski vacuum $|0_M\rangle$, tracing the emergence of thermal behaviour along the alternative paths described in the preceding section. We then repeat the analysis using the Rindler vacuum $|0_{R_1}\rangle$ associated with the wedge  $\mathrm{R_1}$, allowing us to directly contrast the thermal signatures arising from the two different vacuum choices.
 \subsection{Analysis of the four paths via Virasoro Anomaly \label{subsec-4.1}}
 \subsection{Path 1} We obtain the particle content in $R_3$ via path 1.  To investigate the Minkowski Vacuum $|0_M\rangle$, as perceived by a uniformly accelerated Rindler observer via different paths. We are using the Schwarzian derivative to compute the expectation value of the components of the energy-momentum tensor through the Minkowski Vacuum $|0_M\rangle$.
 We have the non-vanishing components are $T_{U_M U_M}$ and $T_{V_M V_M}$. Furthermore, in the Minkowski Vacuum state $|0_M\rangle$,
\begin{equation}
    \langle 0_M|T_{U_M U_M}|0_M\rangle = 0,  \qquad  \langle 0_M|T_{V_M V_M}|0_M\rangle = 0 . \label{Eq:4.1.0.2}
\end{equation}
And the expectation value of the energy-momentum tensor is defined with respect to the Minkowski Vacuum state $|0_M\rangle$,
\begin{equation}
   \langle 0_M|:T_{u u}:|0_M\rangle = -\frac{\hbar}{24\pi}S(U,u), \label{Eq:4.1.0.4}  
\end{equation}
From  Eqs.~(\ref{Eq:2.1.0.3}),~(\ref{Eq:2.1.0.4}),~(\ref{Eq:4.1.0.3}) and ~(\ref{Eq:4.1.0.4}), and after a simple calculation, the expectation value of the energy-momentum tensor in Rindler $\mathrm{R_1}$ gives, 
\begin{equation}
     \langle 0_M|:T_{u_1 u_1}:|0_M\rangle = \frac{\hbar a^2}{48\pi},\label{Eq:4.1.0.5}
\end{equation}
\begin{equation}
    \langle 0_M|:T_{v_1 v_1}:|0_M\rangle = \frac{\hbar a^2}{48\pi},\label{Eq:4.1.0.6}
\end{equation}
Similarly, the the expectation value in $\mathrm{R_3}$ directly with respect to Minkowski Vacuum $|0_M\rangle$ is calculated as:
\begin{equation}
     \langle 0_M|:T_{u_3 u_3}:|0_M\rangle = \frac{\hbar a^2}{48\pi},\label{Eq:4.1.0.7}
\end{equation}
\begin{equation}
    \langle 0_M|:T_{v_3 v_3}:|0_M\rangle = \frac{\hbar a^2}{48\pi},\label{Eq:4.1.0.8}
\end{equation}
We than compute the expectation values in $\mathrm{R_3}$ through the excited state $\mathrm{R_1}$ by the following equations~\cite{Fabbri:2005mw},
\subsection{Path 2\label{subsec-4.2}}
In this section, we again investigate the expectation values of the energy–momentum tensor in the vacuum state as seen from the Rindler wedge $\mathrm{R_1}$, evaluated via paths 2,3,4.

We compute the energy–momentum tensor according to Path-2. In this case, the field is evaluated in $\mathrm{R_3}$, but the vacuum state is taken to be that of the Rindler wedge $\mathrm{R_1}$ and is denoted as $|0_{R_1}\rangle$.

 From Eqs.~(\ref{Eq:2.1.0.11}) and ~(\ref{Eq:4.1.0.3}), we have
 \begin{equation}
  S(u_1, u_3) = -\frac{a^2 e^{a u_3}}{2(1+e^{a u_3})^2} (2 + e^{a u_3}), \label{Eq:4.1.0.16} 
 \end{equation}
 We than compute the expectation values in $\mathrm{R_3}$ through the Vacuum state $\mathrm{R_1}$ by the following equations,
\begin{equation}
  \begin{split}
       \langle 0_{R_1}|:T_{u_3 u_3}:|0_{R_1}\rangle  = \bigg(\frac{\partial_{u_1}}{\partial_{u_3}}\bigg)^2  \langle 0_{R_1}|:T_{u_1 u_1}:|0_{R_1}\rangle\\ - \frac{\hbar}{24\pi} S(u_1,u_3), \label{Eq:4.1.0.17}
  \end{split}
\end{equation}
In the Rindler Vacuum state $\mathrm{R_1}$,
\begin{equation}
    \langle 0_{R_1}|:T_{u_1 u_1}:|0_{R_1}\rangle = \langle 0_{R_1}|:T_{v_1 v_1}:|0_{R_1}\rangle  =0, \label{Eq:4.1.0.18}
\end{equation}
From Eqs.~(\ref{Eq:4.1.0.16}),~(\ref{Eq:4.1.0.17}) and ~(\ref{Eq:4.1.0.18}), we obtain:
\begin{equation}
     \langle 0_{R_1}|:T_{u_3 u_3}:|0_{R_1}\rangle =\frac{\hbar a^2 e^{a u_3}(2+e^{a u_3})}{48\pi (1+e^{a u_3})^2} , \label{Eq:4.1.0.19}
\end{equation}
Eq.~(\ref{Eq:4.1.0.19}) can be evaluated for different limits.
\paragraph{$u_3\to 0$}
\begin{equation*}
   \langle 0_{R_1}|:T_{u_3 u_3}:|0_{R_1}\rangle = \frac{\hbar a^2}{64\pi} 
\end{equation*}
\paragraph{$u_3 \to \infty$}
\begin{equation*}
   \langle 0_{R_1}|:T_{u_3 u_3}:|0_{R_1}\rangle = \frac{\hbar a^2}{48\pi} 
\end{equation*}
\paragraph{$u_3 \to -\infty$}
\begin{equation*}
   \langle 0_{R_1}|:T_{u_3 u_3}:|0_{R_1}\rangle = 0 
\end{equation*}
Similarly, we can calculate,
\begin{equation}
   \langle 0_{R_1}|:T_{v_3 v_3}:|0_{R_1}\rangle =\frac{\hbar a^2(1+ 2e^{a v_3})}{48\pi (1+e^{a v_3})^2} , \label{Eq:4.1.0.20} 
\end{equation}
Eq.~(\ref{Eq:4.1.0.20}) can be evaluated for different limits as above.
\paragraph{$v_3\to 0$}
\begin{equation*}
   \langle 0_{R_1}|:T_{v_3 v_3}:|0_{R_1}\rangle = \frac{\hbar a^2}{64\pi} 
\end{equation*}
\paragraph{$v_3 \to \infty$}
\begin{equation*}
   \langle 0_{R_1}|:T_{v_3 v_3}:|0_{R_1}\rangle =0
\end{equation*}
\paragraph{$v_3 \to -\infty$}
\begin{equation*}
   \langle 0_{R_1}|:T_{v_3 v_3}:|0_{R_1}\rangle =  \frac{\hbar a^2}{48\pi}
\end{equation*}
The relevant limits for our problem are $u_3 \to \infty$ and $v_3 \to \infty$, since we are interested in the particle content near the horizons. We see that it coincides with the answer in path 1. This shows that close to the horizon, there is no difference between the particle content if it is from the Minkowski vacuum or the Rindler vacuum. The distinction is felt when one is far from the horizon. As can be seen in the above equations, the stress tensor approaches zero very far from the horizons, reflecting the fact that the quantum state is the Rindler vacuum. 

\subsection{Path 3} 
We now analyze path 3. We evaluate the expectation values of the energy-momentum tensor in the null-shifted wedge $\mathrm{R_2}$ as shown in Fig.~\ref{Fig:1} with respect to the vacuum state in $\mathrm{R_1}$.
From Eqs.~(\ref{Eq:2.1.1.5}) and  ~(\ref{Eq:4.1.0.3}), we get
\begin{equation}
    S(u_1,u_2) = 0 , \label{Eq:4.1.0.21}
\end{equation}
A simple calculation gives
\begin{equation}
     \langle 0_{R_1}|:T_{u_2 u_2}:|0_{R_1}\rangle = 0,\label{Eq:4.1.0.22}
\end{equation}
This implies there is no flux of left-moving particles, as obtained previously using Bogoliubov coefficients.  We now analyze the other sector.
And similarly, we can calculate,
\begin{equation}
     \langle 0_{R_1}|:T_{v_2 v_2}:|0_{R_1}\rangle = \frac{\hbar a^2(1+ 2e^{a v_2})}{48\pi (1+e^{a v_2})^2},\label{Eq:4.1.0.23}
\end{equation}
Eq.~(\ref{Eq:4.1.0.23}) can be evaluated for different limits as above.
\paragraph{$v_2\to 0$}
\begin{equation*}
   \langle 0_{R_1}|:T_{v_2 v_2}:|0_{R_1}\rangle = \frac{\hbar a^2}{64\pi} 
\end{equation*}
\paragraph{$v_2 \to \infty$}
\begin{equation*}
   \langle 0_{R_1}|:T_{v_2 v_2}:|0_{R_1}\rangle = \frac{\hbar a^2}{24\pi} e^{-a v_2}
\end{equation*}
\paragraph{$v_2 \to -\infty$}
\begin{equation*}
   \langle 0_{R_1}|:T_{v_2 v_2}:|0_{R_1}\rangle =  \frac{\hbar a^2}{48\pi}
\end{equation*}
In the above discussion, we observe that with respect to the Rindler vacuum state, $T_{v_2 v_2}$ has the same flux as thermal radiation in the near-horizon limit. Far from the horizon, the flux goes to zero. We again observe that observers close to the horizon observe a thermal flux of left-moving particles. 

Now we evaluate the expectation values of the energy–momentum tensor in the Rindler wedge $\mathrm{R_3}$ via Path-3, using the  Rindler vacuum defined in $\mathrm{R_1}$, the expectation value of the energy-momentum tensor is
\begin{equation}
\begin{split}
    \langle 0_{R_1}|:T_{u_3 u_3}:|0_{R_1}\rangle  = \bigg(\frac{\partial_{u_1}}{\partial_{u_3}}\bigg)^2  \langle 0_{R_1}|:T_{u_2 u_2}:|0_{R_1}\rangle\\ - \frac{\hbar}{24\pi} S(u_1,u_3), \label{Eq:4.1.0.24}
  \end{split}   
\end{equation}
And,
\begin{equation}
\begin{split}
    \langle 0_{R_1}|:T_{v_3 v_3}:|0_{R_1}\rangle  = \bigg(\frac{\partial_{v_1}}{\partial_{v_3}}\bigg)^2  \langle 0_{R_1}|:T_{v_2 v_2}:|0_{R_1}\rangle\\ - \frac{\hbar}{24\pi} S(v_1,v_3), \label{Eq:4.1.0.25}
  \end{split}   
\end{equation}
 From Eqs.~(\ref{Eq:4.1.0.16}), ~(\ref{Eq:4.1.0.22}) and ~(\ref{Eq:4.1.0.24}), we get
 \begin{equation}
    \langle 0_{R_1}|:T_{u_3 u_3}:|0_{R_1}\rangle =\frac{\hbar a^2 e^{a u_3}(2+e^{a u_3})}{48\pi (1+e^{a u_3})^2},  \label{Eq:4.1.0.26} 
 \end{equation}
 Eq.~(\ref{Eq:4.1.0.26}) is exactly matches with the Eq.~(\ref{Eq:4.1.0.19}). We note that in the limit of $u_3 \to \infty$, the flux matches the thermal flux. So close to the horizon, the stress tensor component indicates thermal distribution of right-moving particles. We now evaluate the particle flux.
 Similarly, by the simple calculations, we get
 \begin{equation}
     \langle 0_{R_1}|:T_{v_3 v_3}:|0_{R_1}\rangle  = \frac{\hbar a^2}{48\pi} \frac{(2 e^{a v_3} +1)}{(e^{a v_3}+1)^4}\big(2 e^{2 a v_3}+2 e^{a v_3} +1\big), \label{Eq:4.1.0.27}
 \end{equation}
 Eq.~(\ref{Eq:4.1.0.27}) can be evaluated for different limits as above.
\paragraph{$v_3\to 0$}
\begin{equation*}
   \langle 0_{R_1}|:T_{v_3 v_3}:|0_{R_1}\rangle = \frac{5\hbar a^2}{256\pi} 
\end{equation*}
\paragraph{$v_3 \to \infty$}
\begin{equation*}
   \langle 0_{R_1}|:T_{v_3 v_3}:|0_{R_1}\rangle = \frac{\hbar a^2}{12\pi} e^{-a v_3}
\end{equation*}
\paragraph{$v_3 \to -\infty$}
\begin{equation*}
   \langle 0_{R_1}|:T_{v_3 v_3}:|0_{R_1}\rangle =  \frac{\hbar a^2}{48\pi}
\end{equation*}

\subsection{Path 4} 
Finally, we evaluate the expectation values of the energy-momentum tensor in the Rindler wedge $\mathrm{R_3}$ along Path-4 as in the above calculations. In this path, the field configuration is mapped according to $\mathrm{R_1} \longrightarrow \mathrm{R_4} \longrightarrow \mathrm{R_3}$,
where the intermediate wedge $\mathrm{R_4}$ plays the role opposite to that of $\mathrm{R_2}$ in Path-3. Specifically, while in $\mathrm{R_2}$ we found 
$\langle T_{u_2 u_2} \rangle = 0$ and a nonvanishing $\langle T_{v_2 v_2} \rangle$, the corresponding expectation values in $\mathrm{R_4}$ are simply interchanged, with $\langle T_{v_4 v_4} \rangle = 0$ and $\langle T_{u_4 u_4} \rangle$ taking the same functional form as $\langle T_{v_2 v_2} \rangle$. When this $\mathrm{R_4}$  is propagated into $\mathrm{R_3}$, the derivative factors and Schwarzian contributions appearing in the $\mathrm{R_4} \to \mathrm{R_3}$ transformation reproduce exactly the same expressions as those obtained along Path-3. As a result, the final expectation value of the stress-energy components,$\langle 0_{R_1}|: T_{u_3 u_3}: |0_{R_1} \rangle, \qquad \langle 0_{R_1}|: T_{v_3 v_3}: |0_{R_1} \rangle$, computed via Path-4 coincide identically with the corresponding components derived along Path-3. This confirms that both paths lead to the same stress-energy tensor in $\mathrm{R_3}$, demonstrating the internal consistency and path independence of the setup.
\par

\section{Conclusions and Discussion}
In the article, we showed that there are multiple paths that yield a thermal distribution of particles in a given Rindler wedge, when we focus on particle content close to the horizon. For observers close to the horizons, we see that all four paths yield the same thermal density for both left-moving and right-moving particles. The various paths, based on the analysis in the article, are: a) Minkowski spacetime with scalar field in vacuum state,  b) Rindler wedge 1 $R_1$ with the scalar field in Rindler vacuum, c) Rindler wedge $R_2$ with with a thermal flux of left moving particles d) Rindler wedge $R_4$ with a thermal flux of right moving particles.
Here are a few questions that can be asked based on the above observations. We note that there is an inherent degeneracy when defining a given path.  The degeneracy arises from varying the ``distances" along the null and spatial shifts. There are various superset wedges (a few spatially shifted, and a few null shifted with respect to the wedge $R_3$ that yield the same  particle content close to the horizon in $R_3$. 

Firstly, can one define a notion of entropy associated with a Rindler wedge with the observed thermal density of particles? 
How to use this degeneracy to consistently define a notion of information entropy is a question we leave for future consideration.

An important consequence of the result could be for Hawking radiation and the evaporation of a black hole. The qualitative arguments are presented below.  The black hole evaporation is modeled as follows. The black hole emits a flux of radiation at a temperature inversely proportional to the mass $M$ of the black hole (when we assume the cosmological constant to be zero). The total mass, therefore, decreases to $M-dM$. The black hole now radiates at a higher temperature because its mass has decreased. The black hole continues to lose mass until it completely evaporates. If we understand this process in discrete steps. The exterior of the black hole with mass $M$ is a subset of the new exterior of a black hole with mass $M-dM$. The near-horizon geometry of the black hole with mass $M$ resembles a Rindler wedge, complete with features like thermal flux of particles. We name it $R_2$. Now, when we go to the next step in Hawking evaporation, we consider a superset of $R_2$, which is again a Rindler wedge in the near-horizon geometry. So the main question is ``Which superset does it decay into?" Is it another wedge similar to $R_2$ with again a thermal flux of particles? The current models of Hawking radiation assume that the superset again has a thermal flux of radiation. The work presented in this article suggests a different possibility. The superset could be a wedge $R_1$ in vacuum. If this is the case, Hawking radiation will temporarily cease when the black hole mass reaches $M-dM$, since the superset is in a vacuum. Of course, it can be argued that a Rindler vacuum in the current setting is similar to the Boulware vacuum and might not be physically realizable. This decay could be unstable and has to be explored rigorously. If this decay mode is allowed, the continuous evaporation of a black hole might be punctuated by a series of pauses and bursts of radiation. 

\appendix
\section{Calculation of Bogoliubov coefficient \label{Apn1}}
Bogoliubov coefficient $\alpha_{21}^{*}(\Omega\omega)$ is written as\begin{equation}
\alpha_{21}^{*}(\Omega,\omega) = \frac{1}{2\;\pi} \sqrt{\frac{\Omega}{\omega}}\int_{-\infty}^{\infty} dv_2\; e^{\;i\;\Omega\; v_2}\; e^{\;-i\;\omega \;v_1} \label{Apn1.1}
\end{equation}
Substituting Eq.~(\ref{Eq:2.1.1.10})  in the  ~(\ref{Apn1.1}),we have
\begin{equation}
   \alpha^{*}_{21}(\Omega,\omega)= \frac{1}{2\;\pi}\sqrt{\frac{\Omega}{\omega}}\int_{-\infty}^\infty dv_2 \;e^{\;i\;\Omega\; v_{2}} e^{ \frac{\;-i\;\omega \;ln[e^{\;a \;v_2}+ 1]}{a}} \label{Apn1.2}  
\end{equation}
Hence, ~(\ref{Apn1.2}) can be written as
\begin{equation}
  \alpha^{*}_{21}(\Omega,\omega)= \frac{1}{2\pi}\sqrt{\frac{\Omega}{\omega}}\int_{-\infty}^\infty dv_2\; e^{\;i\;\Omega\; v_{2}} ( e^{\;a \;v_2}+ 1)^{-\frac{i\;\omega}{a}} \label{Apn1.3} 
\end{equation}
By substituting$e^{\;a \;v_2}+ 1=h\implies h-1=e^{\;a \;v_2}$ and also $dv_2= \frac{dh}{a( h-1)}$ in the above equation.Hence ~(\ref{Apn1.3}) reduces to,
\begin{equation}
  \alpha^{*}_{21}(\Omega,\omega)= \frac{1}{2\pi}\sqrt{\frac{\Omega}{\omega}}\int_{2 a \Delta_{1}}^\infty dh\; ( h-1)^{\frac{i\;\Omega}{a}-1} h^{-\frac{i\;\omega}{a}} \label{Apn1.4}
\end{equation}
Again, substituting $ h-2 \;a \;\Delta_{1}=t\implies h= t+2\; a\; \Delta_{1}$ and also $dh=dt$ in ~(\ref{Apn1.4}) and hence we can write,
\begin{equation}
  \alpha^{*}_{21}(\Omega,\omega)= \frac{1}{2\;\pi \;a}\sqrt{\frac{\Omega}{\omega}}\int_{0}^\infty dt \;t^{\frac{i\;\Omega}{a}-1}\; (t+1)^{-\frac{i\;\omega}{a}} \label{Apn1.5}
\end{equation}
The integral in ~(\ref{Apn1.5}) can be evaluated from Mathematica and therefore can be written as
\begin{equation}
 \alpha^{*}_{21}(\Omega,\omega)= \frac{1}{2\;\pi\; a}\sqrt{\frac{\Omega}{\omega}}\frac{2^{\frac{-i\;\omega+i\;\Omega}{a}}\;\Gamma[\frac{i\;\Omega}{a}]\;\Gamma[\frac{i\;\omega-i\;\Omega}{a}]}{\Gamma[\frac{i\;\omega}{a}]}\label{Apn1.6}
\end{equation}
From  Eq.~(\ref{Eq:3.2.0.9}), Bogoliubov coefficient $\beta_{21}^{*}(\Omega\omega)$ is written as
 \begin{equation}
     \beta^{*}_{21}(\Omega,\omega)=-\frac{1}{2\;\pi} \sqrt{\frac{\Omega}{\omega}}\int_{-\infty}^\infty dv_2\;e^{\;i\;\Omega\; v_{2}}\; e^{\;i\;\omega\; v_{1}}\label{Apn1.7}
  \end{equation}
As the above calculation ~(\ref{Apn1.7}) can be written as
  \begin{equation}
  \beta^{*}_{21}(\Omega,\omega)= -\frac{1}{2\;\pi\; a}\sqrt{\frac{\Omega}{\omega}}\int_{0}^\infty dt \;t^{\frac{i\;\Omega}{a}-1}\; (t+1)^{\frac{i\;\omega}{a}} \label{Apn1.8}
\end{equation}
By the integration technique  ~(\ref{Apn1.8}) can be written as
 \begin{equation}
\beta^{*}_{21}(\Omega'\omega)= -\frac{1}{2\;\pi \;a}\sqrt{\frac{\Omega}{\omega}}\frac{2^{\frac{i\;(\omega+\Omega)}{a}}\;\Gamma[\frac{i\;\Omega}{a}]\;\Gamma[\frac{-i\;(\omega+\Omega)}{a}]}{\Gamma[\frac{-i\;\omega}{a}]}, \label{Apn1.9}
\end{equation}

\section{Particle number density  \label{Apn2}}
We solve the integral ~(\ref{Eq:3.2.0.26}) in this appendix.
Let;
\begin{equation}
    I = \int_{0}^\infty d\Omega\;\frac{sinh\;(\frac{\pi\;\Omega}{a})}{(\nu+\Omega)\;sinh\;[\frac{\pi(\nu+\Omega)}{a}]} ,\label{Apn2.1}
\end{equation}
Let; $\frac{\pi\;(\nu+\Omega)}{a}= z \implies \frac{\pi\;\Omega}{a}= z-\frac{\pi\;\nu}{a}$,
Also; $d\Omega= \frac{a\;dz}{\pi}$

Substituting these values in the Integral ~(\ref{Apn2.1}),we obtain,
\begin{equation}
   I = \int_{\frac{\pi\;\nu}{a}}^\infty \frac{dz}{z}\frac{sinh\;(z-{\frac{\pi\nu}{a}})}{sinh\;z},\label{Apn2.2}
\end{equation}
After some algebra, The Integral  ~(\ref{Apn2.2}) can be written as
\begin{equation}
   I = \int_{\frac{\pi\;\nu}{a}}^\infty \frac{dz}{z}\;\big[cosh\;(\frac{\pi\;\nu}{a})-coth\;z\; sinh\;(\frac{\pi\;\nu}{a})\big],  \label{Apn2.3}
\end{equation}
By substituting The Integral ~(\ref{Apn2.3}) into Eq.~(\ref{Eq:3.2.0.16}), we get
\begin{equation}
     \langle\hat{N}(\nu)\rangle = \frac{cosh\;(\frac{\pi\;\nu}{a})}{4\;\pi \;a\; sinh\;(\frac{\pi\;\nu}{a})}\int_{\frac{\pi\;\nu}{a}}^\infty \frac{dz}{z} - \frac{1}{4\;\pi\; a }\int_{\frac{\pi\;\nu}{a}}^\infty \frac{dz}{z}\;coth\;z ,\label{Apn2.4}
\end{equation}
 Eq.~(\ref{Apn2.4}) further split into two integrals,
\begin{equation}
    \langle\hat{N}(\nu)\rangle =  I_{1}-I_{2}, \label{Apn2.5}
\end{equation}
Here;
\begin{equation}
   I_{1} =  \frac{cosh\;(\frac{\pi\;\nu}{a})}{2 \;sinh\;(\frac{\pi\;\nu}{a})}\int_{\frac{\pi\;\nu}{a}}^\infty \frac{1}{2\;\pi \;a}\frac{dz}{z}, \label{Apn2.6}
\end{equation}
The left-hand side of Eq.~(\ref{Apn2.5}) gives the dimensionless number of particles. So, $I_1$ and $I_2$ must also be dimensionless. To keep the units consistent in  Eq.~(\ref{Apn2.6}), the factor $\frac{1}{2\pi a}$is included inside the integral.
The integral $\int_{\frac{\pi\nu}{a}}^\infty \frac{1}{2\pi a} \frac{dz}{z}$ exhibits a logarithmic ultraviolet (UV) divergence arising from contributions of high-frequency modes. This type of divergence is characteristic of quantum field theories in curved spacetime and often reflects the accumulation of an infinite density of states near horizons or boundaries. In such contexts, divergences are frequently formalized via expressions like $\delta(0)$—interpreted as an infinite volume or mode density. $\delta(0)$ emerges from continuous mode normalization and is directly linked to spatial volume~\cite[Sec.2.3]{TongQFT}.
 So the Integral  ~(\ref{Apn2.6}) can be written after simplification;
\begin{equation}
       I_{1}  = \frac{1}{2}\frac{e^{\frac{2\;\pi\;\nu}{a}}}{e^{\frac{2\;\pi\;\nu}{a}}-1}\;\delta(0) +  \frac{1}{2}\frac{1}{e^{\frac{2\;\pi\;\nu}{a}}-1}\;\delta(0), \label{Apn2.7}
\end{equation}
Now;
\begin{equation}
   I_{2} =  \frac{1}{4\;\pi\; a }\int_{\frac{\pi\;\nu}{a}}^\infty \frac{dz}{z}\;coth\;z ,\label{Apn2.8}
\end{equation}
\begin{figure}[!ht]
	\centering
	\includegraphics[scale=0.5]{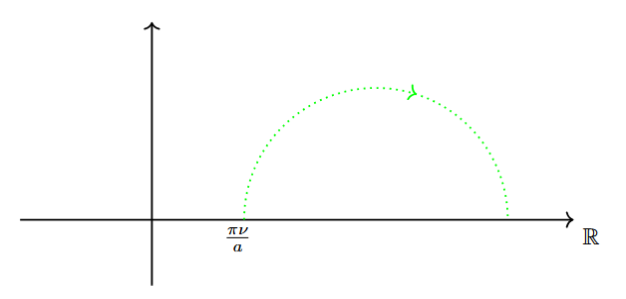}   
	\caption{Complex plane  \label{Fig:4}}
\end{figure}
The integral  ~(\ref{Apn2.8}) has a simple pole at $z= n\;\pi\; i$ and also a singularity at $z=0$, as shown in the contour of Fig.~\ref{Fig:4}

By the expansion of $cothz$,
\begin{equation*}
    cothz = \frac{1}{z}+\frac{z}{3}-\frac{z^3}{45}+ \cdots
\end{equation*}
Explicitly stretching the contour in Fig.~\ref{Fig:4} by $R\to \infty$ to $R\to -\infty$.
\begin{figure}[!ht]
	\centering
	\includegraphics[scale=0.5]{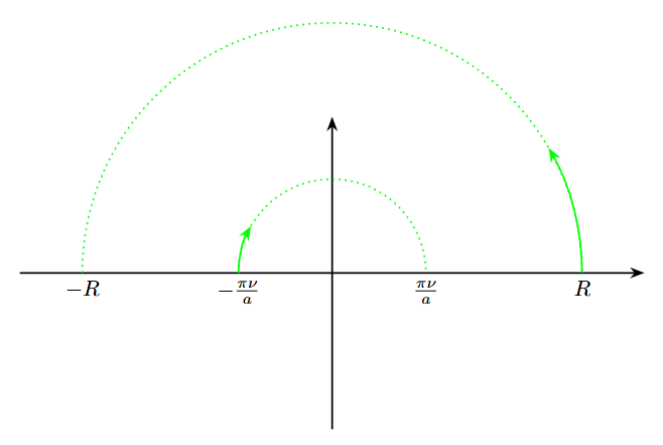}
	\caption{Complex plane  \label{Fig:5}}
\end{figure}
Now, by the expansion of $coth\;z$ at the pole, $z=n\;\pi\; i$,
\begin{equation}
    cothz = \frac{1}{z-n\;\pi \;i}+Reg. part ,\label{Apn2.9}
\end{equation}
Hence, the residue is
\begin{equation*}
   \begin{split}
        Residue = \lim_{z\to n\;\pi\; i }\; (z-n\;\pi\; i)\frac{cothz}{z} = \lim_{z \to n\pi i }\frac{1}{z}\\
    =\frac{1}{n\pi i}
   \end{split}
\end{equation*}
Since;$\frac{1}{\pi\; i}\sum_{n=1}^{\infty} \frac{1}{n}$
Now;
\begin{equation}
  \sum_{n=1}^{\infty} \frac{1}{n} = 1+\frac{1}{2}+\frac{1}{3}+\frac{1}{4}+\cdots=\infty,\label{Apn2.10}
\end{equation}

Applying the Residue Theorem to a contour, as shown in Fig.~\ref{Fig:5},
\begin{equation}
    \begin{split}
        \int_{-R}^{-\frac{\pi\;\nu}{a}}f(z)\;dz + \int_{\frac{\pi\;\nu}{a}}^{R}f(z)\;dz + \int_{0}^{\pi}f(R\;e^{i\;\theta})\;i\;R\;e^{\;i\;\theta}\;d\theta+\\
    \int_{\pi}^{0}f(\frac{\pi\;\nu}{a}e^{\;i\;\theta})\;i\;\frac{\pi\;\nu}{a}\;e^{\;i\;\theta}d\theta = 2\pi i(\frac{1}{\pi i})(\infty),\label{Apn2.11}
    \end{split}
\end{equation}
Where; $f(z)= \frac{1}{z}\;cothz$
By taking the limit $R\to \infty$ and after further simplification of Eq. ~(\ref{Apn2.11}) reduces to be;
\begin{equation}
    2\int_{\frac{\pi\nu}{a}}^{\infty}\frac{1}{z}\;coth\;z\;dz +i \int_{\pi}^{0}\frac{\pi\;\nu}{a}\;coth\;(\frac{\pi\;\nu}{a}e^{\;i\;\theta})\;d\theta =\infty , \label{Apn2.12}
\end{equation}
Since; $i \;\int_{\pi}^{0}\frac{\pi\;\nu}{a}\;coth\;(\frac{\pi\;\nu}{a}e^{\;i\;\theta})\;d\theta= -i\frac{\pi^2\;\nu}{a}\; cot\;h(\frac{\pi\;\nu}{a}\;e^{\;i\;\theta})$, which has a finite value.Hence,  ~(\ref{Apn2.12}) reduced to be,
\begin{equation}
    \int_{\frac{\pi\nu}{a}}^{\infty}\frac{1}{z}\;coth\;z\;dz = \infty ,\label{Apn2.13}
\end{equation}
In quantum field theory, integrals that reduce to infinity due to a simple pole can be interpreted in a distributional sense as Dirac delta functions~\cite{Peskin:1995ev, Srednicki:2007qs, Weinberg:1995mt}. Hence, Eq. ~(\ref{Apn2.13}) can be written as
\begin{equation}
    \int_{\frac{\pi\;\nu}{a}}^{\infty}\frac{1}{2\;\pi\; a}\frac{1}{z}coth\;z\;dz = \delta(0) ,\label{Apn2.14}
\end{equation}
To maintain dimensional consistency, the factor $\frac{1}{2\pi a}$ is included in the integral.
So, from  Eq. ~(\ref{Apn2.8}) and  Eq. ~(\ref{Apn2.14}),
\begin{equation}
    I_{2} =  \frac{1}{2} \; \delta(0) ,\label{Apn2.15}
\end{equation}
Substituting  ~(\ref{Apn2.7}) and   ~(\ref{Apn2.15}) into  ~(\ref{Apn2.5}) and after some algebra, we obtain 
\begin{equation}
     \langle\hat{N}(\nu)\rangle =  \frac{\delta(0)}{e^{\frac{\;2\;\pi\;\nu}{a}}-1}, \label{Apn2.16}
\end{equation}
Reflecting the familiar Bose–Einstein distribution at the Unruh temperature, with \(\delta(0)\) interpreted as the formal infinite spatial volume factor.

\section{Bogoliubov Transformation for two observes in Rindler spacetime\label{Subsec-3.6}}
 \begin{figure}[!ht]
	\centering
	\includegraphics[scale=0.7]{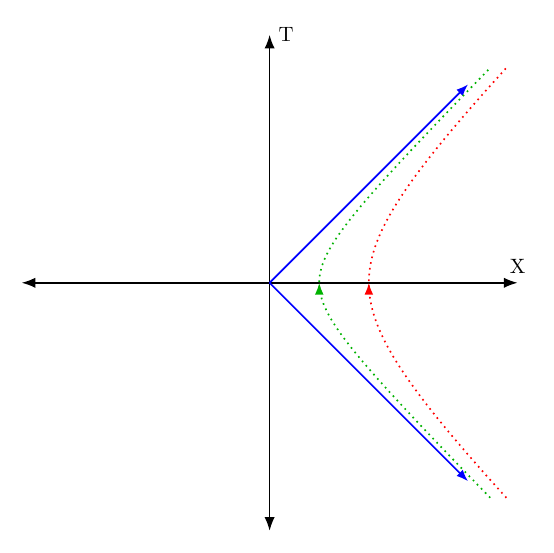}   
	\caption{Rindler spacetime with observer having acceleration a and g with green and red respectively.\label{Fig:6}}
\end{figure}
We now consider the relation between two Rindler observers accelerating with proper accelerations $a$ and $g$, respectively. The coordinate transformation from Minkowski to Rindler coordinates with acceleration $a$ is given by:
\begin{equation}
	T = \frac{e^{a \;x_1}}{a} \;\sinh\;(a \;t_1), \label{Eq:3.6.0.1}
\end{equation}
\begin{equation}
	X = \frac{e^{a\; x_1}}{a}\; \cosh(a\; t_1) .\label{Eq:3.6.0.2}
\end{equation}
which in light-cone coordinates takes the form:
\begin{equation}
    U = T-X = -\frac{e^{-a \;u_1}}{a}, 
    \label{Eq:3.6.0.3}
\end{equation}
\begin{equation}
    V = T+X = \frac{e^{a\; v_1}}{a}. 
    \label{Eq:3.6.0.4}
\end{equation}
Similarly, for Rindler coordinates with acceleration $g$, we have:
 \begin{equation}
	T = \frac{e^{g\; x_2}}{g} \;\sinh\;(g \;t_2), \label{Eq:3.6.0.5}
\end{equation}

\begin{equation}
	X = \frac{e^{g\; x_2}}{g}\; \cosh(g\; t_2) ,\label{Eq:3.6.0.6}
\end{equation}
and in light-cone form:
\begin{equation}
    U = T-X = -\frac{e^{-g \;u_2}}{g}, 
    \label{Eq:3.6.0.7}
\end{equation}
\begin{equation}
    V = T+X = \frac{e^{g\; v_2}}{g}, 
    \label{Eq:3.6.0.8}
\end{equation}
By equating the light-cone coordinates $U$ and $ V$  in the two charts, we find the transformations between the null coordinates:
\begin{equation}
    u_1 = \frac{-1}{a}\;ln\bigg(\frac{a}{g}\bigg)+\frac{g}{a}\;u_2,\label{Eq:3.6.0.9}
\end{equation}

\begin{equation}
   v_1 = \frac{1}{a}\;ln\bigg(\frac{a}{g}\bigg)+\frac{g}{a}\;v_2,   \label{Eq:3.6.0.10}
\end{equation}
Using these relations, we evaluate the Bogoliubov coefficients for right-moving modes via:
\begin{equation}
 \overrightarrow{\alpha^{*}}_{\Omega,\omega}= \frac{1}{2\pi}\sqrt{\frac{\Omega}{\omega}}\int_{-\infty}^\infty du_2\; e^{i\;\Omega\; u_2}\; e^{-i\;\omega \;u_1} \label{Eq:3.6.0.11}
  \end{equation} 
  \begin{equation}
     \overrightarrow{\beta^{*}}_{\Omega\omega}= -\frac{1}{2\pi}\sqrt{\frac{\Omega}{\omega}}\int_{-\infty}^\infty du_2 \;e^{i\;\Omega\; u_2}\; e^{i\;\omega\; u_1}, \label{Eq:3.6.0.12}
  \end{equation} 
  Substituting Eq.~\eqref{Eq:3.6.0.9} into Eq.~\eqref{Eq:3.6.0.11}, we obtain:
  \begin{equation}
 \overrightarrow{\alpha^{*}}_{\Omega,\omega} = \frac{1}{2\pi}\sqrt{\frac{\Omega}{\omega}}\bigg(\frac{a}{g}\bigg)^{\frac{i\omega}{a}}\int_{-\infty}^\infty du_2\; e^{{i\; u_2}\;\bigg(\Omega-\frac{\omega g}{a}\bigg)},\label{Eq:3.6.0.13}
  \end{equation}
  By the integration technique, we obtained the following result.
  \begin{equation}
     \overrightarrow{\alpha^{*}}_{\Omega,\omega}= \sqrt{\frac{\Omega}{\omega}}\;\bigg(\frac{a}{g}\bigg)^{\frac{i\;\omega}{a}}\delta\bigg(\Omega-\frac{\omega\; g}{a}\bigg), \label{Eq:3.6.0.14}
  \end{equation}
  Similarly, the $\beta^*$-coefficient becomes
   \begin{equation}
 \overrightarrow{\beta^{*}}_{\Omega,\omega}= -\frac{1}{2\pi}\;\sqrt{\frac{\Omega}{\omega}}\;\bigg(\frac{a}{g}\bigg)^{\frac{-i\;\omega}{a}}\int_{-\infty}^\infty du_2 \;e^{{i\; u_2}\;\bigg(\Omega+\frac{\omega\; g}{a}\bigg)},\label{Eq:3.6.0.15}
  \end{equation} 
  Hence, by integrating Eq.~(\ref{Eq:3.6.0.15}), we get
  \begin{equation}
     \overrightarrow{\beta^{*}}_{\Omega,\omega}= 0 ,\label{Eq:3.6.0.16} 
  \end{equation}
  Following a similar analysis for left-moving modes, we obtain:
  \begin{equation}
  \overleftarrow{\alpha^{*}}_{\Omega,\omega}  = \sqrt{\frac{\Omega}{\omega}}\bigg(\frac{a}{g}\bigg)^{\frac{-i\;\omega}{a}}\delta\bigg(\Omega-\frac{\omega\; g}{a}\bigg) ,\label{Eq:3.6.0.17}
  \end{equation}

  \begin{equation}
      \overleftarrow{\beta^{*}}_{\Omega,\omega}= 0, \label{Eq:3.6.0.18} 
  \end{equation}
  \noindent
Thus, both left- and right-moving Bogoliubov coefficients  $\beta_{\Omega\omega} $ vanish. This indicates that the Rindler vacuum defined with acceleration $a$ appears as a vacuum to observers accelerating with $g$, i.e., no particle creation occurs between these frames.

\begin{acknowledgments}
The authors thank the BITS Pilani Hyderabad campus for providing the necessary infrastructure for this research. RK thanks Amita for her valuable discussions on Appendix B.
\end{acknowledgments}
\bibliography{Ref.bib}
\end{document}